\documentclass[10pt,conference]{IEEEtran}
\IEEEoverridecommandlockouts
\usepackage{cite}
\usepackage{amsmath,amssymb,amsfonts}
\usepackage{graphicx}
\usepackage{textcomp}
\usepackage{xcolor}
\usepackage{booktabs} % For formal tables
\usepackage{colortbl}
\usepackage{amsmath}
\usepackage{algorithm}
\usepackage{algpseudocode}
\usepackage[normalem]{ulem} %sout
\usepackage{xspace}
\usepackage{multirow} % Fancy tables
\usepackage{balance} % Reference balancing
\usepackage{fancyvrb} % Verbose mode in footnotes
\usepackage[export]{adjustbox} % Aligning differently-sized subfigures
\usepackage[caption=false]{subfig}
\usepackage{url}
\usepackage{hyperref}
\usepackage{xcolor}
\usepackage{array}
\usepackage{fontawesome}

\usepackage{mdframed}
\mdfsetup{skipabove=0.5\topskip,skipbelow=0.5\topskip}

\usepackage{enumitem}
\setlist[itemize]{leftmargin=*,noitemsep,topsep=0pt}
\setlist[enumerate]{leftmargin=*}

\newenvironment{smalldescription}{
   \setlength{\topsep}{0pt}
   \setlength{\partopsep}{0pt}
   \setlength{\parskip}{0pt}
   \begin{description}[style=unboxed]
   \setlength{\leftmargin}{1in}
   \setlength{\parsep}{0pt}
   \setlength{\parskip}{0pt}
   \setlength{\itemsep}{0pt}
   }
   {\end{description}}

\usepackage{cleveref}
\crefformat{section}{\S#2#1#3}
\crefname{figure}{Figure}{Figures}
\crefname{table}{Table}{Tables}

\usepackage{todonotes}
\usepackage{soul}

\newcommand{\ie}{\textit{i.e.,}\xspace}
\newcommand{\eg}{\textit{e.g.,}\xspace}
\newcommand{\etal}{\textit{et al.}\xspace}

\newcommand{\REDOS}{ReDoS\xspace}

\newcommand{\SurveyNumValidResponses}{158\xspace}

\newcommand{\TotalPercKnowREDOS}{38\%\xspace}

\newcommand{\JamesSurveyNumberResponses}{121\xspace}

\def\BibTeX{{\rm B\kern-.05em{\sc i\kern-.025em b}\kern-.08em
    T\kern-.1667em\lower.7ex\hbox{E}\kern-.125emX}}

\usepackage{pifont}% http://ctan.org/pkg/pifont

\begin{document}
\VerbatimFootnotes % Allow \verb in footnotes

\title{Regexes are Hard: Decision-making, Difficulties, and Risks in Programming Regular Expressions}
 \author{\IEEEauthorblockN{Louis G. Michael IV}
 \IEEEauthorblockA{Virginia Tech \\ louism@vt.edu }
 \and
 \IEEEauthorblockN{James Donohue}
 \IEEEauthorblockA{University of Bradford \\ j.donohue@bradford.ac.uk }
 \and
 \IEEEauthorblockN{James C. Davis}
 \IEEEauthorblockA{Virginia Tech \\ davisjam@vt.edu }
 \and
 \IEEEauthorblockN{Dongyoon Lee}
 \IEEEauthorblockA{Stony Brook University \&\\Virginia Tech \\ dongyoon@cs.stonybrook.edu }
 \and
 \IEEEauthorblockN{Francisco Servant}
 \IEEEauthorblockA{Virginia Tech \\ fservant@vt.edu }
 }
%\author{\IEEEauthorblockN{Anonymous Regular Expression Investigators}}
\maketitle

\begin{abstract}
Regular expressions (regexes) are a powerful mechanism for solving string-matching problems.
They are supported by all modern programming languages, and have been estimated to appear in more than a third of Python and JavaScript projects.
Yet existing studies have focused mostly on one aspect of regex programming: readability.
We know little about how developers perceive and program regexes, nor the difficulties that they face.

%While regex are highly expressive, they are also often perceived to be highly complex and hard to read. While existing studies have focused on improving the readability of regex, little is known about any other difficulties that developers face when programming with regex.
In this paper, we provide the first study of the regex development cycle, with a focus on 
(1) how developers make decisions throughout the process,
(2) what difficulties they face, and
(3) how aware they are about serious risks involved in programming regexes.
We took a mixed-methods approach, surveying 279 professional developers from a diversity of backgrounds (including top tech firms) for a high-level perspective, and interviewing 17 developers to learn the details about the difficulties that they face and the solutions that they prefer.

In brief, regexes are hard.
Not only are they hard to read, our participants said
that they are hard to search for,
hard to validate,
and hard to document.
%Developers also reported cascading impacts of poor readability, lack of universal portability, and struggling with overall problem comprehension.
They are also hard to master: the majority of our studied developers were unaware of critical security risks that can occur when using regexes, and those who knew of the risks did not deal with them in effective manners.
Our findings provide multiple implications for future work, including semantic regex search engines for regex reuse and improved input generators for regex validation.

\end{abstract}

\begin{IEEEkeywords}
regular expressions, developer process, qualitative research
\end{IEEEkeywords}

\section{Introduction}

%\FS{Clean-up: consistently capitalize section titles.} I belive this is fixed now
%\FS{Clean-up: consistently say ``reuse'' or ``re-use''.} Went with reuse

%\JD{In my rewrite I collapsed the details of our ``marriage'' methodology. It felt too detailed to give in the introduction.}
%\DY{This looks fine.}

% \FS{TODO: Strengthen motivation and expected implications. Revisit this section answering the questions:
% What problem is the paper addressing?
% Why is the problem important?
% What are the existing solutions to this problem?
% What is the proposed solution in this paper?
% Why is the proposed solution different or more promising than the existing solutions?
% How did they evaluate that the proposed solution in fact improves the state of the art?
% To what extent does the proposed solution improve the state of the art?
% How are the results useful to anybody else? / ``who cares?''
% }
% \DY{Cosmetic: do not have white space and tilde together before {\textbackslash}cite. they make two whitespaces.}
% \DY{Three ideas in one paragraph: (1) regex is popular; (2) regex is hard; and (3) regex has little studied. Putting (2)(3) separated from (1) and making them richer (as FS suggested) would motivate the paper better.}

% Everybody uses regexes
Regular expressions (regexes) are a text processing tool baked into all modern programming languages as well as popular tools like text editors~\cite{SublimeSearchAndReplace,VSCodeBasicEditing}.
Developers frequently incorporate regexes into their software.
Estimates suggest that more than a third of JavaScript and Python projects include at least one regex~\cite{Chapman2016ExploringPython,Davis2018EcosystemREDOS}.
% Transition.
It is therefore surprising that we know so little about how developers interact with such a widely-used technology.

%They are not just widely supported but also prevalent having been identified in 42.2\% of randomly-selected GitHub Python repositories ~\cite{Chapman2016ExploringPython} and \cite{Davis2018EcosystemREDOS} found that over 30\% of npm and pypi modules contain regexes.
% and have been of increasing concern as a security risk ~\cite{Davis2018EcosystemREDOS, Staicu2018REDOS, Davis2018NodeCure}.

% We are focusing on the technology, not on the developers.
Existing investigations into regexes have mainly focused on the source code and the technology rather than on the person using it.
For example, other studies have delved into
regex feature usage~\cite{Chapman2016ExploringPython},
regex evolution~\cite{WangExploringEvolution}, 
and worst-case regex behavior~\cite{Davis2018EcosystemREDOS}.
%\DY{This is a good sentence to have if it is true. But I feel a bit of over-claiming or naive over-simplification: e.g., We know people don't test regexes well.}
%\MM{I walked it back a little from we know about}
%\JD{@DY I disagree. We know about the ARTIFACTS -- developer test suites don't test regexes well -- but nothing about the PROCESS. For example some of our findings indicate that traditional test suite analysis will not capture all of the tests that go into validating regexes. I don't think we are overclaiming.}
%\DY{Well. I originally wanted to limit our claim to PROCESS (as you mentioned). I can see that ``how software developers actually work with regexes'' is the process. Disregard my comments.}
However almost nothing is known about how software developers actually work with regexes.
Without understanding what real developers
  think about regexes,
  how they go about creating them,
  what difficulties they face,
  and how they manage risks,
we can only guess at what aspects can be improved.

%This popularity is despite the fact that regexes are frequently considered difficult to write ~\cite{Chapman2017ExploringComprehension, Spishak2012AExpressions}.
%Regex usage trends have been studied \cite{Hodovan2010RegularWeb, Chapman2016ExploringPython}, researchers have performed empirical evaluations of their readability \cite{Chapman2017RegexComprehension}, and even looked at how they change over time \cite{WangExploringEvolution}, but almost nothing is known about how actual professional developers, the people who author regexes with real world impact, actually go about the process of creating them.
%Without understanding what real developers think about regex, how they go about creating them, what difficulties they face, and what risks they may encounter we can only guess at what can be improved.
% \DY{I suggest the following sentence becomes the topic sentence of the next paragraph. A single-sentence goal statement.}

% We did it.
We present the first large-scale qualitative study of the decision-making, difficulties, and risks developers face when they program with regexes.
We surveyed 279 professional developers about their regex practices, and support our findings by interviewing 17 professional developers to illuminate our findings.
%This work is a marriage of an investigation from a single company background and a broad professional screening interacting with \TotalCombinedResponses developers from with a range of skills and backgrounds.
% What did we learn?
Through our investigation, we gained insight into developer perceptions and decision-making regarding regexes.
%Developers view regexes as a powerful tool with a steep learning curve.
We report on the decisions developers make when working on regexes: to apply a regex or not; to write or reuse a regex; to identify test input; and whether the regex is correct. 
%  shedding light on the difficulties that developers feel when wrestling through problems. 
We shed light on the difficulties that developers feel when wrestling through problems,
  ranging from mundane syntax problems to complex reasoning about the relative importance of false positives and false negatives in the pattern-matching problem space.
We learn about how developers work to handle these obstacles,
  the tactics they employ and the tools they use.
We examine the risks of regex programming, with the surprising result that many developers are unaware of the portability and security problems associated with regexes.
% These two populations were asked similar survey and interview questions but with an evolving focus.
% To gain the fullest picture of developer perception, decision making, challenges, how they deal with these challenges, and awareness of risks regarding regex the team conducted a series of distinct probes asking similar related but not identical question to each population.   
% \DY{I suggest we have a summary finding paragraph. Interesting, counter-intuitive findings, etc.}

Our core contributions are as follows:
\begin{itemize}
    % Data collection
    \item In order to provide as rich a glimpse of regex practices as possible, we conduct and analyze large-scale surveys of 279 professional software developers from a diversity of companies, including several major tech firms to understand their regex practices. We further our findings by interviewing 17 developers.
    % Data analysis:  
    \item We synthesize this mixed-methods data to better understand the regex programming process: the decisions developers make, the difficulties developers face and how they handle them, and the degree of awareness around the risks of regex programming.
    \item We discuss the myriad implications of our findings, proposing a wide range of directions for future research grounded in real-world practice.
\end{itemize}

\section{Background and Related Work} \label{sec:background}
% \subsection{Regex}
%  Regex are a way of expressing a formal language and were first used by Stephen Kleene. ~\cite{Kleene1951RepresentationAutomata}. They 
% \MM{Intro what is regex and it has worst case}
%  \MM{Provide a definition for a string matching problem here}

\subsection{Regex Programming and Risks}

%\JD{Do we want to give a general description of the regex programming process?}
%\MM{I would shy no since we kind of impose one in the methodology}
%\DY{I suggest we give a general desp here. This is a good place to explain what it means by ``assessing the problem'' (i.e., defining the set of strings to match and not to match)}
%\JD{@DY I agree}
Regexes are a string matching tool, used to identify a generalized subsequence of characters within a string.
Generally, the process that ends with the inclusion of a regex in code written by a professional developer takes four overall steps. 
First, the developer identifies --- or is tasked with --- a string matching problem that he or she assesses for its suitability to be solved using a regex.
Next, developers will move to compose a regex, evaluating the feasibility of reuse, and then, either author a regex from scratch or reuse one, with or without modification.
Then, having arrived at a regex that the developer hopes solves their problem, they will validate it.
If the regex passes validation, developers will document the regex and integrate it into their project. 
If the regex fails validation, developers will attempt to recompose it.

%\JD{Separate subsection for these? Put them right up front? Preceded by some general summary of the regex programming process?}
%\DY{I am fine with putting this upfront, given that it has already demoted ReDoS discussion.}
Software developers face two major risks when programming regexes: portability and performance.
Many regex dialects have emerged over the years~\cite{friedl2006mastering}, with divergent syntaxes and semantics.
Developers therefore face \textit{portability problems} during regex programming, with the risk that the regex that they compose or reuse will be executed in a dialect other than the one they anticipate, with unanticipated behavior (\eg syntax errors or unexpected match behavior)\cite{Davis2019WhyExpressions}.

Developers also face \textit{performance problems leading to security risks} due to the polynomial or exponential worst-case time complexity of regex matches in most regex engines~\cite{Cox2007RegexAlgorithms}
These super-linear regexes can expose applications to Regular expression Denial of Service (\REDOS) vulnerabilities~\cite{Crosby2003REDOS,Crosby2003AlgorithmicComplexityAttacks}, which have been reported on dozens of major websites~\cite{Staicu2018REDOS}, hundreds of major JavaScript projects~\cite{Davis2018NodeCure}, and thousands of JavaScript, Python, and Java projects~\cite{Wustholz2017Rexploiter,Davis2018EcosystemREDOS}.
Any software developers who write client-facing regexes face the risk of regex performance problems and \REDOS security vulnerabilities.

\subsection{Empirical Regex Research}

Prior research regarding regexes has been predominantly quantitative, examining regexes in their role as a software artifact.
Researchers have empirically examined
regex reuse~\cite{Hodovan2010RegularWeb,Davis2019WhyExpressions},
regex test coverage~\cite{Wang2018RegexTestCoverage},
regex evolution~\cite{Wang2019ExploringEvolution},
regex repair~\cite{Davis2018EcosystemREDOS},
and
regex generalizability~\cite{davis2019generalizability}.
Others have proposed tools for regex programming, \eg
input generation~\cite{Larson2016GeneratingExpressions,Veanes2010Rex,Arcaini2017MutRex,Shen2018ReScueGeneticRegexChecker,Moller2010Brics},
linters~\cite{Larson2018AutomaticCheckingOfRegexes},
and
type checking~\cite{Spishak2012AExpressions}.
% Respond to artifact-style
Although regexes are interesting artifacts, they also represent hours of developer effort that bear qualitative investigation.
Our developer-focused investigation of regex programming is orthogonal to quantitative research that treats regexes as a software artifact.
% Respond to tool-style
Our qualitative efforts will inform the development of new tools that address the problems developers actually face, maximizing the potential impact of regex tool research.

%\JD{The next paragraph may be a bit strong.}

Only two studies have explored the developer side of regex programming, with an emphasis on composition and comprehension.
% In the only survey of professional developers in the literature, 
Chapman and Stolee~\cite{Chapman2016RegexUsageInPythonApps} asked 18 professional developers how often and in what contexts they use regexes.
And in a laboratory setting, Chapman \etal~\cite{Chapman2017RegexComprehension} performed a fine-grained study on whether their subjects preferred one regex ``synonym'' over another (\eg equivalent patterns that use character classes \verb|/[ab]/| or disjunctions \verb</a|b/<).
% Motivate our work as important -- let's get out of the laboratory
% \DY{I suggest we drop the following sentence (to avoid any potential offensive tone). I think our point is that existing two works are small-scale, focused on one or two aspects. I think that msg has been delivered. Jumping directly to ``Our approach is to cast ...'' looks fine to me.}
% Counterintutively, they reported that participants improved in comprehension tasks as regexes grew longer --- directly contradicting most of our interview participants, and suggesting that talking to developers \textit{in situ} may yield more practical insights than laboratory experiments.
Our approach is to cast our net broadly, hearing from hundreds of developers from diverse backgrounds to understand coarse-grained issues surrounding process, difficulties, and risks.

\subsection{Developer Perceptions, Practices and Information Needs}

We are not the first to apply qualitative methods to shed light on software engineering practices.
Prior work in this vein ranges from general engineering perceptions~\cite{Li2015WhatEngineer} to specific practices such as code review~\cite{Bacchelli2013ExpectationsReview}.
The standard approach is to survey~\cite{Bacchelli2013ExpectationsReview, Gousios2015WorkPerspective, Ernst2015MeasureDebt, Buse2012InformationAnalytics} or interview or observe~\cite{Li2015WhatEngineer, Bacchelli2013ExpectationsReview, Ernst2015MeasureDebt, Fannoun2019TowardsPractice, Ko2007InformationTeams} developers who are exposed to the topic of interest.
Some studies have also considered interaction artifacts~\cite{Breu2010InformationReports, Bacchelli2013ExpectationsReview}.

Our study is the first investigation of developer regex programming practices in this spirit.
We are
  the first to survey developers on many of these regex topics at all,
  the first to survey developers on regexes at scale,
  and the first to conduct regex-focused developer interviews of any kind.

\begin{figure*}[htbp]
\centerline{\includegraphics[width=2.0\columnwidth]{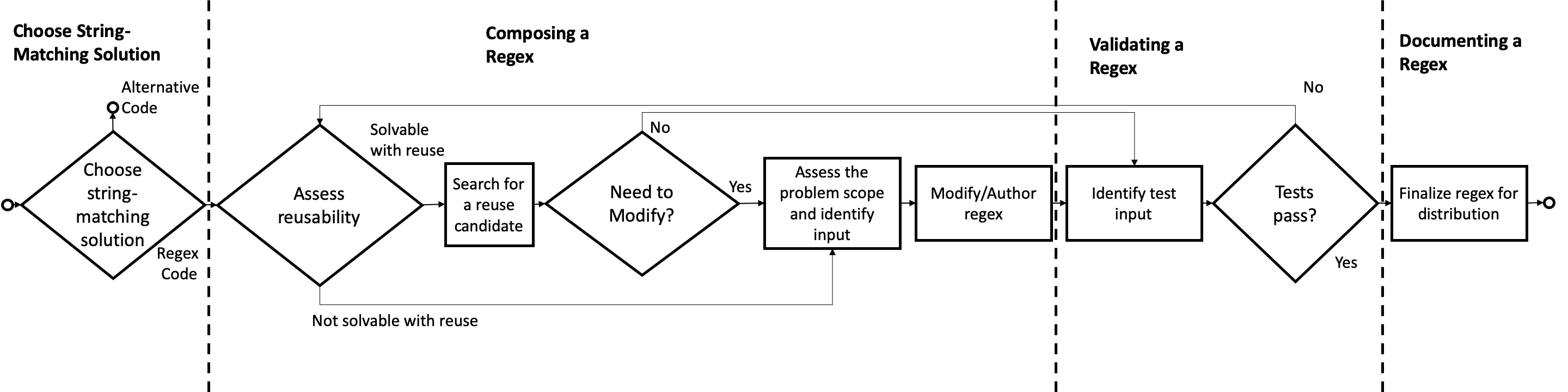}}
\caption{Stages of regex programming, with four major decision points (diamonds). This outline is what we used to frame the further investigation into developers' decision-making processes and difficulties.}
\label{fig:fullProcess}
\end{figure*}

\section{Research Questions}
\label{sec:RQs}

In this study, we focus on understanding core human aspects of regex programming: how developers make their decisions, what difficulties they face, and whether they are aware of dangerous risks.
Understanding these aspects of regex programming will motivate impactful new lines of research targeting the specific problems that professional software developers face.
To this end, we study the following research questions:
\begin{smalldescription}

\item[RQ1:] What perceptions do developers have about the value and difficulty of regexes?
\item[RQ2:] What influences developer decisions when programming regexes?
\item[RQ3:] What do developers find difficult about programming regexes? 
\item[RQ4:] How do developers handle those difficulties in programming regexes? 
\item[RQ5:] Are developers aware of portability and security risks when programming regexes?

\end{smalldescription}

To support our study of these research questions, we devised a general framework for the regex programming process, depicted in~\cref{fig:fullProcess}.
We adapted this framework from the general software engineering methodology of defining requirements, writing, validating, and deploying~\cite{Pressman2010SoftwareApproachb}, and introduced a reuse stage based on our intuition that regexes are a kind of ``function'', complex enough to be reused like other software.
We do not claim that our framework is exhaustive, but we believe it captures the crucial regex programming decisions that developers make.
Using this framework, we were able to focus our methodology on developers' decision-making, challenges, and handling-mechanisms for each of the decisions in~\cref{fig:fullProcess}.
%that is well understood by professional developers --- but with a specific focus on regex programming.
%We expanded on this general idea by including the specifics of regex reuse, following our intuition that regexes would be reused much the same way as other code is reused. 

%, which we pose is a set of decisions that are important in the process.
%We designed our framework to target diverse contexts in the study of our research questions, \ie to ask developers about multiple stages of their process of programming regexes.
%We ask developers about their decision-making, challenges, and handling-mechanisms for the decisions that we display in \cref{fig:fullProcess} --- which we pose is a set of decisions that are important in the process.

\section{Research Method}\label{section:Methodolgy}

\textit{Overall method and considerations.}
% overall method
We followed a mixed qualitative and quantitative approach~\cite{creswell2017research}.
We used a qualitative approach to answer most of our questions, but for RQ1 and RQ5 we also asked some quantitative questions.
We emphasized a qualitative approach with the goal of \emph{discovery}: to identify an exhaustive set of answers to our research questions, as well as understand the details and context behind them.

% why
Since discovery was our goal, we wanted to maximize the number of professional developers whose perspectives we heard through surveys and interviews.
Of course, one major difficulty in qualitative software engineering research is persuading enough (busy, highly-paid) subjects to participate to give weight to findings.
We therefore prepared two distinct pairs of surveys and interview protocols, with different emphases roughly on the left and right halves of the process framing~\cref{fig:fullProcess}.
This allowed us to reach a diverse population of software developers and to ask a diverse set of questions, while also keeping the survey-taking time to a reasonable 17-minute median and the interview times to roughly 30 minutes.
%We wanted to administer surveys and interviews along two lines general regex practices and regex reuse practices, and devised appropriate survey instruments and interview protocols.
%In order to obtain sufficient survey and interview subjects, we divided each pair of instruments into four pieces: two distinct survey instruments and two distinct interview protocols, halving the 

%With this goal of maximizing the discovery of answers for our research questions, we performed two surveys and complemented them with two sets of follow-up semi-structured interviews.
% \FS{median survey-taking time?}
% \MM{17.5 min}

% surveys
%Thus, one survey mainly focused on general regex practices, and a second mainly focused on questions about regex reuse.

% open questions
\textit{Survey design.}
% pilots
We designed our surveys by first having discussions with professional software developers, and following best practices in survey design~\cite{Kitchenham2008PersonalSurveys,Siegmund2014MeasuringExperience}.
Then, we refined the design of each survey through internal iteration, followed by several pilot administrations each.
%for six pilot for the first survey and two pilot rounds for the second one.

Both of our final survey instruments included free response questions about the four stages of regex programming that we set out to study (see \cref{sec:RQs}).
In these open questions, we asked developers about their thought process when making the decisions that we described in \cref{sec:RQs}, and we asked them what they found difficult in programming with regexes.
We also asked them to describe the mechanisms that they follow to handle those difficulties.
% connection to RQs
We used these responses to answer our research questions RQ2, RQ3 and RQ4.

% mc questions
Our surveys also included multiple-choice questions about developers' perceptions about regexes in general, which we used to answer RQ1.
Then, we also included multiple-choice questions asking about developers' awareness of portability and ReDoS risks, and free response questions asking them how they prevent such risks.
We used these responses for RQ5.

Finally, our second survey also included three additional questions about regex reuse, the results of which are discussed in a previous publication that focuses specifically on the topic of regex reuse~\cite{Davis2019WhyExpressions}.

\textit{Survey deployment.}
% recruitment
We deployed our surveys after obtaining approval from our institution's ethics board, following a two-pronged strategy to maximize the diversity of respondents.

We deployed our first survey internally in a large international media company. 
% recruitment-detailed
%We used convenience sampling to recruit staff who worked with regular expressions as part of their jobs inside the company. 
We sought participants through an internal advertising campaign and by asking senior engineering staff to promote the survey.

We deployed the second survey at software companies of various sizes. %and on Internet forums.
We used snowball sampling~\cite{Biernacki1981SnowballSampling,Sadler2010ResearchStrategy}, contacting professional developers of our acquaintance who work at tens of different software companies, including top Fortune 500 companies, and asking them to take the survey and propagate it to their colleagues.
To further increase the diversity of responses, we also posted the survey on popular Internet message boards that are frequented by software developers~\cite{HackerNews,Reddit}).
We compensated legitimate responses with cookies for the first survey and a \$5 Amazon gift card for the second one.

% Responses
We obtained survey responses from \JamesSurveyNumberResponses developers for our first survey and \SurveyNumValidResponses developers for our second one.
% : more than 6 years for our first survey, 
The median respondent had more than 6 years of professional experience in the first survey and 3-5 years in the second one.
% \FS{Add cool sentence about why we asked for demographics at the end.}

% saturation

% interviews
\textit{Interview design and deployment.}
The final question in our surveys was a request for permission to conduct a follow-up interview.
We contacted all survey respondents who were willing to be interviewed, and were able to schedule interviews with 17 of them.
% \FS{How many interviews did we conduct?}
Following common practice to learn more about a phenomenon~\cite{weiss1995learning}, our interview protocol was semi-structured~\cite{lindlof2017qualitative}.
We developed \emph{interview guides} with general topics and questions to be covered instead of an exact set of questions.
%This approach is often used to obtain more insights about a phenomenon \cite{weiss1995learning}.
We focused our interview guides on the decisions that developers make, the difficulties that they face, and the ways in which they handle them when programming regexes.
We also asked for clarification on details of the regex programming process, hinted at by our survey results.
We compensated interview participants from the second interview population with a \$25 Amazon gift card.
The first set of interview participants were not compensated.

\textit{Data analysis.}
% data analysis
  % coding, other steps...
    % factors for decisions
    % difficulties
    % handling mechanisms
% \FS{I saw this paper saying ```To analyze the more than 60 hours of interviews and 388,000 words of transcripts...''' Can we say something like that?}
% \MM{We did something like 10 hours between the two of us}
We analyzed the free response questions in our surveys and the transcriptions of our interviews using \emph{open coding} \cite{l2001qualitative} (also used in \emph{grounded theory} \cite{adolph2011using}).
% This should be a section that discusses our actual coding practices since that was asked for by reviewers
For our second survey, one author of the paper read all the responses for a question to identify codes into a code book.
Afterwards, the author reread and coded all the responses.
Then, a different author of this paper used the code book to perform their own coding of units.
Finally, both sets of codes were used to reach agreement.
Due to organizational privacy and confidentiality requirements, the first survey was analyzed by one author, coding free responses and then repeating the process several weeks later, using the codes from the other survey. 
The results were then compared with the original codes to resolve discrepancies.
Finally, we organized the resulting codes according to the research question that their corresponding quotes answered.
We report them together with exemplary quotes in \cref{section:RQ1,section:DevProcess,section:Challenges,section:Security}.
% Each iteration of coding focused on grouping individual responses into related categories and eventually grouping these categories into more unifying ideas.
% This methodology most closely aligns with open coding followed by axial coding but the disjoint nature of the two surveys made it difficult to follow grounded theory closely. 
Our survey instruments and interview protocols are available for replication at \url{http://doi.org/10.5281/zenodo.3424069} \cite{louis_g_michael_iv_2019_3424069}.

\section{RQ1: What perceptions do developers have about the value and difficulty of regexes?}\label{section:RQ1}
\begin{figure}[htbp]
\centerline{\includegraphics[width=1.0\columnwidth]{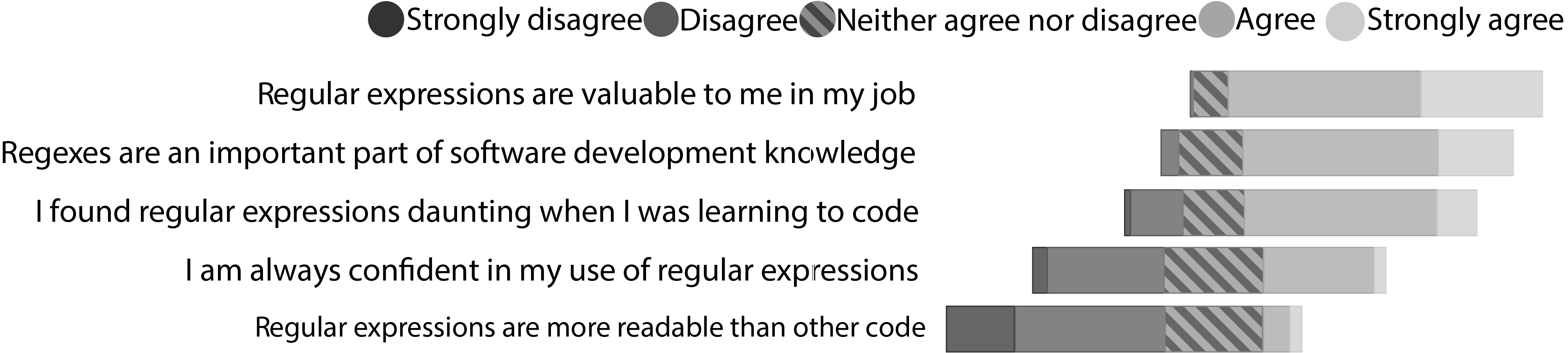}}
\caption{We asked developers about their perceptions of regexes, in regards to: value to their job (most agreed that they are valuable), and as software developer knowledge (most agreed that they are important).
Regarding the difficulty of regexes, most developers: agreed that they were daunting when learning to code, did not show strong confidence in their regex usage (overall neutral response), and disagreed that regexes are more readable than other code.
%\FS{Please, align this graph.}
% \DY{In the legend, let's not use ``...''}
}
% \JDo{Strictly speaking, owing to the wording these do not show that the developers view regexes as less readable etc.}
\label{fig:regex-opinions}
\end{figure}

We initially wanted to understand whether regexes are a technology that developers value and benefit from,
as well as to learn developers' general perceptions about their difficulty.
We found that \textbf{developers view regexes as a valuable technology, but one that is also difficult} (\cref{fig:regex-opinions}).

To understand whether developers believe that regexes are valuable, our first survey asked participants if they agreed with the statement ``Regular expressions are valuable to me in my job''.
Respondents were nearly unanimous: 88\% of developers agreed that regexes were valuable to their job.
%Only 1 out of the \JamesSurveyNumberResponses developers surveyed disagreed with this statement and, 
Developers also generally agreed that ``Regular expressions are an important part of software development knowledge''.
% \DY{I see that RQ1 has not been updated. But just for reference, we want to share two msgs in RQ1: (1) regexes are valuable, yet (2) regexes are considered to be hard.}

But our participants also described regexes as difficult.
65\% of respondents agreed that they ``found regular expressions daunting when\ldots learning to code'';
most developers did not always feel confident in their usage of regexes;
and most developers (70\%) disagreed with the statement that ``regular expressions are more readable than other code''. 

These initial findings give strong motivation to pursue regex research in general,
and prompted us to investigate our subsequent research questions to characterize the particular ways in which developers find regexes difficult.
% While regex are frequently refereed to as difficult in literature to our knowledge this claim is much more an intuition then a supported claim.
% We surveyed ACME developers on this and more generally on their perceptions of regex and regex usage.  
% They preserved regex as both more error prone then non regex code and are less readable, while also being a valuable job skill ~\cref{fig:regex-opinions}. 
% In interviews developers built on this idea discussing regex as a tool that is not frequently preferred but is also at times the only method available or the only tool with the capabilities of solving some problems.
% ``I'd rather use non regex string matching where possible, but it's not always powerful and flexible enough to, \ldots [when using regex] deciding to write my own or not is that I'm just going to get it wrong.''
% ACME developers also indicated that they believed regex are an important part of software development knowledge but were ``daunting'' when first learning.
% Ended up using a different interview \MM{7:30 AM Monday interview quotes about how they are hard to read and not maintainable, good for ad hoc usage because that is where they are required}

% \section{Decision Making Process}\label{section:DevProcess}
\section{RQ2: What Influences Developer Decisions when Programming Regexes?}\label{section:DevProcess}
% \MM{We talked about chaning the wording of this on Friday at 10AM}

{
\renewcommand{\arraystretch}{1.2}
\begin{table*}
%\footnotesize
\small
\centering
%\raggedright
% \caption{Developer-reported Decision Factors when Programming Regexes}
\caption{Developer-reported Decision Factors when Programming Regexes and their Influence on Decision Outcome}
\begin{tabular}{llll}
\toprule
% \textbf{Stage}											& \textbf{Decision}													& \textbf{RQ2: Factors}														\\
% \toprule
% Choosing a Solution								& 1) Using regex vs. using non-regex code			& Problem complexity, single choice, readability				\\
% \arrayrulecolor{gray}
% \hline
% \multirow{3}{*}{Composing a Regex}		& 1) Writing regex vs reusing regex						& Problem commonality, reliability, time savings				\\
% 																& 2) Which regex should I pick for reuse?				& Best understanding, feature usage, length						\\
% 																& 3) Match too much vs match too little?				& Problem domain																\\
                      
% \hline
% \multirow{2}{*}{Validating a Regex}			& 1) Am I confident this regex is correct?				& Choosing input or having sample data, result recipient	\\
% 																& 2) Am I confident this reused regex is correct?	& Trust of reuse																	\\
% \hline
% \multirow{1}{*}{Documenting a Regex}	& 1) How much documentation is required?			& Personal opinion, complexity											\\
\textbf{Stage}							& \textbf{Decision}												& \textbf{RQ2: Factors}		& \textbf{Influence on Outcome} \\
\toprule
\multirow{3}{*}{Choosing a Solution}	& \multirow{3}{*}{1) Using regex vs. using alternative code}	& Problem complexity 		& Medium-complexity \faicon{arrow-right} Regex \\
										&																& Readability				& Both outcomes \\
										&																& Single choice				& Regex \\
\arrayrulecolor{black}
\hline
\multirow{7}{*}{Composing a Regex}		& \multirow{3}{*}{1) Writing regex vs reusing regex}			& Problem commonality		& Common problem \faicon{arrow-right} Reuse \\
										&																& Reliability 				& Reuse \\
										&																& Time savings				& Both outcomes \\
\arrayrulecolor{gray}
\cline{2-4}
										& \multirow{3}{*}{2) Which regex should I pick for reuse?}		& Best understanding		& Simpler regexes preferred \\
										&																& Feature usage				& Fewer preferred \\
										&																& Length					& Shorter preferred \\
\arrayrulecolor{gray}
\cline{2-4}
										& 3) Match too much vs match too little?						& Problem domain			& Domain-dependent \\
                      
\arrayrulecolor{black}
\hline
\multirow{3}{*}{Validating a Regex}		& \multirow{2}{*}{1) Am I confident this regex is correct?}		& Sample data availability	& Available \faicon{arrow-right} High confidence \\
										&																& Result recipient			& Internal use \faicon{arrow-right} Shallow testing \\
\arrayrulecolor{gray}
\cline{2-4}
										& 2) Am I confident this reused regex is correct?				& Trust of reuse			& Both outcomes \\
\arrayrulecolor{black}
\hline
\multirow{2}{*}{Documenting a Regex}	& \multirow{2}{*}{1) How much documentation is required?}		& Personal opinion 			& Both outcomes \\
										&																& Complexity				& High \faicon{arrow-right} Documentation \\
\arrayrulecolor{black}
\bottomrule
\end{tabular}
\label{table:DecisonFactors}
%\vspace{-0.6cm}
\end{table*}
}

One of the primary results of our work was an understanding of how developers make the decisions involved in programming regexes.
We report it in the following subsections.

\subsection{Choosing a String-matching Solution}

We asked developers how they made the decision of which string matching solution to use, asking specifically about using a regex vs. writing alternative code

% \FS{I think that we should define ``string-matching problem'' early in the paper. In background or methodology.}

\subsubsection{Using Regex vs. Using Alternative Code}

% \FS{For each decision, let's mention the literal question that we asked for this purpose, and say that we report on the answers to this question and the clarifications about this question in the interviews.}

% \FS{Consider clarifying what questions are we reporting answers from.
% For this subsection, I expect answers will come from 
% ``Imagine a simple string validation problem that could be solved either using a regex or other built-in language/platform features. In general, which would you prefer?'', 
% ``Please can you explain the reason for your previous answer?''
% If we are using quotes from answers to other questions, we should say so.
% }

% \FS{I further analyzed your argumentation to distill the factors that developers considered for making this decision, and making the argumentation (paragraphs) revolve around that. I am keeping the previous version commented out below, so that you can see what I did here.}

% based on complexity
Developers reported making this decision based mostly on the \textbf{perceived complexity} of the problem.
Developers perceive regexes as well suited for solving ``Goldilocks'' problems, neither too simple nor too complex.
For simple problems, simple string APIs were preferred.
As one interviewee said, \emph{``if there's a string function that says the prefix should be this, I would prefer that over a regex \ldots it's simpler to understand''}
And when the problems are too complex, a survey respondent cautioned that regexes are also not a good solution:
\emph{``If a regex is complex enough that it's `too complex' to write from scratch, it's probably also too complex a problem to solve with a regex''}

Another factor that developers considered when deciding to use a regex or its alternatives is their \textbf{perceived readability}.
But developers disagreed on how readable regexes were, with some inclined towards and others away from using them.
One participant said \emph{``I stay away from tedious string parsing and splitting, and I see regular expressions as a tool to aid in conveying what you are trying to do.''}, while other teams discourage regexes instead: \emph{``With code review you want readable code and regexes are often not readable so they become less commonly used''}.

%were split in that this factor may incline them either way, \ie either towards using regexes or towards using alternative code.
%\eg .
%In contrast, other developers considered that, despite being hard to read, regexes may still be more readable and intuitive than other alternatives, such as multiple string functions, for some problems.  
%In an interview one participant framed this by saying:
% \emph{``I'll always try to write the code that is the most readable to the developer \ldots stay away tedious, string parsing and splitting, and I see regular expressions as a tool to aid in conveying what you are trying to do.''}

% \DY{In the meeting, I suggest that we put the following to the end and treat it lightly.}
% based on availability
Developers also mentioned that sometimes \textbf{a single choice is available}.
% \MM{Better GREP cite jamie offered, it may get moved to discussion}
% Examples of such situations are those when regexes are the syntax supported by the tools being used, such as  grep \cite{grep} or the \emph{find} (Ctrl+F) functionality in a text editor.
For example, some built-in language and third-party APIs require developers to supply regexes to solve string matching problems.
A common example is search tools:
%In some situations, the context will determine regexes as the only option to solve the string-matching problem \eg
% \FS{Again, elaborate more}
% \MM{Found a really good quote but I am worried that it is too long. This is from transcript (7) the 5/6 7:30 AM interview}
\emph{``You find regular expressions and globs in search tools all over the place\ldots in those cases, it's not really a choice.''}

\subsection{Composing a Regex}

When developers decide to solve their string-matching problem with a regex,
we asked them how they make two decisions while composing regexes.
First, we asked how they decided to write from scratch vs. re-using a regex.
And when they opt to reuse, we asked them how they select reuse candidate(s).
Developers also volunteered a decision that we had not initially considered, namely determining the relative merits of overly liberal or conservative matching (\ie too much or too little?).
We updated our interview protocol to investigate this decision.

\subsubsection{Writing Regex vs. Reusing Regex}
\label{sec:decisionRegexvs.Non}

Our participants often said they would reuse when they believed they were trying to solve a \textbf{common problem}.
As one survey respondent said, \emph{``If it's a common regex like various form fields I would reuse a regex, but for a more company/business use case specific requirement I would write a custom regex''}.

Many respondents preferred to reuse regexes where possible to \textbf{improve reliability}, believing that regexes from a \textbf{trusted reuse source} would provide higher quality or better testing.
%This notion was frequently accompanied by related factors like gauging the \textbf{trustworthiness of the source}, would provide a \textbf{higher quality} and or a \textbf{better tested} regex then one they may produce on their own.
For example, one interviewee said \emph{``[A highly up voted Stack Overflow regex] is more likely to be right and account for edge cases''}
%than  I come up with on the spot''}
% \MM{This quote is from the 5/7 8:30 PM interview at minute mark 5}
The specifics of what constituted a trusted source varied.
Some developers relied on a private team resource like a shared regex file, while others trusted highly up-voted Stack Overflow posts.
% This is very intuitive since it makes little sense to start looking for something that you do not believe exists. 
% This meant that many developers noted that if they viewed their use case as something that was domain or business specific they would not try to reuse.
% \MM{This is from thought process question}

% \FS{Consider clarifying what questions are we reporting answers from.
% For this subsection, I expect answers will come from 
% ``Why have you in the past decided to re-use regular expressions as opposed to writing them yourself from scratch?'', 
% ``Can you tell us more about your thought process when you decide to reuse a regex vs writing it from scratch?''
% If we are using quotes from answers to other questions, we should say so.
% }

% Developers indicated that after deciding that after they wanted to solve a problem the next decision was whether or not to reuse a regex. 
% \begin{figure}[htbp]
% \centerline{\includegraphics[width=1.0\columnwidth]{figs/MischaSurvey/whyReuseFixed.png}}
% \caption{When surveyed about what factors make them choose to reuse aggregate developers most commonly selected that they thought their use case was common.}
% \label{fig:whyReuse}
% \end{figure}

% The effort of performing the search for a valid reuse candidate is only undertaken when there is hope that something will be found.
% \MM{I don't know if we can say this super strongly or not}

Another factor that developers considered when deciding whether to reuse was \textbf{saving time}, but they disagreed about the more efficient strategy.
%Additionally developers discussed \textbf{saving time}.
Some favored reuse (\emph{``Similar to writing code, finding a working example and adapting it is faster''}), while others said that \emph{``Writing from scratch often requires less time than searching for a suitable one.''}

\subsubsection{Which Regex Should I Pick for Reuse?}
Developers who opt to reuse must often choose from multiple candidate regexes.
Our participants said they preferred simpler regexes when given the choice, but measured complexity in different ways.
%reuse multiple options are available and in these cases developers described having some factors they considered to select a regex. 
One interviewee showed a preference for the \textbf{fewer special characters}, saying \emph{``I just try and pick the one I have the most understanding of \ldots the one with the fewest special characters''}.
%This developer associated understanding with \textbf{feature usage} but other developers mentioned understanding as being linked to \textbf{length}.
%Length was a factor considered for more than just understanding however, also being considered an important factor intrinsically. 
Another emphasized \textbf{length}.
% \emph{``It's also governed by if something's ridiculously complex and I looked at another answer and it's half the length I'm going to go with that one\ldots I feel like that's more concise''}
\emph{``[If one] answer is half the length I'm going to go with that one.''}.

%\emph{``Sometime you find a couple of different answers \ldots I just try and pick the one I have the most understanding of, which is usually the one with the fewest special characters''}
%This developer associated understanding with \textbf{feature usage} but other developers mentioned understanding as being linked to \textbf{length}.
%Length was a factor considered for more than just understanding however, also being considered an important factor intrinsically. 
%An interviewed developer expressed this when they said,
%% \emph{``It's also governed by if something's ridiculously complex and I looked at another answer and it's half the length I'm going to go with that one\ldots I feel like that's more concise''}
%\emph{``It's also governed by if something's ridiculously complex and I looked at another answer and it's half the length I'm going to go with that one.''}
%when discussing choosing from several reuse candidates. 

\subsubsection{Match too much vs. match too little?}

We did not initially anticipate this decision, but added it to our interview protocol based on information volunteered in an early interview.
When composing a regex, developers discussed needing to have an understanding of the context in which their regex would be used.
In some situations, they preferred their regex to be overly liberal, matching too much, while in others they wanted to be conservative, matching too little.
They said \emph{``what might be tricky is deciding whether or not you want to match it too much or match too little''}, and pointed to validation as a \textbf{particular context} where matching too little (false negatives) was preferable to matching too much:
\emph{``I'd much rather match too little than too much [to avoid introducing] garbage data''}.

\subsection{Validating a Regex} \label{sec:RQ2-validating}

%\DY{I am not sure why we are not explictly asking ``how developers validate the correctness of a regex?'' Especially, I feel weird asking Am I ``confident'' that this regex correct?}
We asked developers about two aspects of validation: how they decide whether a regex is correct, and whether their process changes when they are re-using a regex rather than writing their own.
%how they make two decisions when composing regexes: deciding whether a regex is correct, (1) generally, and (2) specifically for reused regexes.

\subsubsection{Is This Regex Correct?}

%\DY{how choosing input is different from having sample data?}
Our participants' confidence was often tied to having \textbf{comprehensive sample input} for their regex.
\emph{``I'm usually pretty confident about them\ldots we have a pretty much an unlimited ... sample pool of things I can use to test.''}
On the other hand, a participant described editing a colleague's old regex as difficult because they did not perfectly understand the input space.
%sample data   Whereas most times I'm editing a colleague's regular expression. I don't tend to have [sample data].''}

%Many developers when discussing validating \textbf{choosing input or having sample data} to test with as major difference in validating.
%One interviewed developer spoke to this when they were discussing the difficulties when they would write a regex vs they would have to edit a colleagues regex.  
%\DY{I don't get the point of introducing a quote for mine vs. others regex. Are we answering what makes regex testing hard?}
%\emph{``I'm working on data, I can plug it into regex101 and plug in my sample data. Whereas most times I'm editing a colleague's regular expression. I don't tend to have [sample data].''}
%\DY{I am not sure why we focus on ``confidence''.}
%Having large amounts of sample data also let developers have more confidence that their solution was accurate.

%\DY{I am not sure what's the point of reporting a bad practice (of not testing software). What's the message? Are we saying regexes are often not tested?}
%Other developers did not do much if any formalized validation and don't worry too much about exact correctness. 
%From an interview where the developer was asked about validation difficulties, 
% \emph{``I just kind of eyeball it, if it seems to work, we just run with it. Then somebody down the line, we'll probably test the edge cases by passing in, something that shouldn't be in there and  it'll be caught at that point''}

In \textbf{non-customer-facing contexts}, some participants had a lower standard for correctness.
They said things like
\emph{``I just kind of eyeball it \ldots somebody \ldots will probably test the edge cases''}, usually when the recipient of the data was a team member rather than a customer.

\subsubsection{Is this Reused Regex Correct?}

% \FS{Consider clarifying what questions are we reporting answers from.
% For this subsection, I expect answers will come from 
% ``Does your validation strategy change when re-using a regex instead of writing it from scratch?''
% If we are using quotes from answers to other questions, we should say so.
% }

% \begin{figure}[htbp]
% \centerline{\includegraphics[width=1.0\columnwidth]{figs/MischaSurvey/validateChange.png}}
% \caption{Example of a figure caption.}
% \label{fig:validateChange}
% \end{figure}
% When aggregate developers where surveyed on if their validation practices change between when the authored from scratch vs when they chose to reuse they predominantly said no that their validation was the same independent of source, ~\cref{fig:validateChange}.
% Those that responded yes were asked how their process changed and the responses generally fell into two categories, more confident in reused so less validation required, and more skeptical of reused regex prompting more thorough evaluation. 
% This stark dichotomy was very interesting.

%\DY{It is not clear to me why we have introduced this second question. Do we have a hypothesis that validation would be different for reused regexes? I think we should motivate why we are asking this question and seeking an answer. I don't see much difference here. Mentioning a mixed trust and distrust of reusing looks to be the key, but I am not sure if this is interesting to discuss.}

Our participants were split on whether and how to change their validation strategy for reused regexes.
Some developers viewed reused regexes as better tested or more and as a result \textbf{did not work to validate as thoroughly} as when they wrote from scratch, saying things like \emph{``I'll usually trust re-using an expression more \ldots [and] skip [some validation phases]''}.
But others treated reused regexes cautiously, \emph{``I am aware of the security implications of using something from public sources''}.
%Despite the possible correctness risks of accidentally re-using regexes across regex dialects (\cref{sec:background}),

%One developer wrote \emph when surveyed on how they change their validation process for reused regexes.
%On the other hand some developers had a natural \textbf{distrust of reusing} and would work harder to more thoroughly validate expressions.  A different developer responded to the same question: \emph{``When using external code I am aware of security implications of using something from public sources, so I double check what I copy.''}

\subsection{Documenting a regex}

\subsubsection{How Much Documentation is Required?}
Perhaps related to the differing opinions on the readability of regexes that we identified earlier, interviewees \textbf{disagreed} on the extent to which \textbf{regex documentation was required}.
Many felt that a well-written regex would not require documentation: \emph{``I would say that \ldots most people would consider them self documenting,''}.
Others thought that the amount of documentation depended on regex complexity, \eg
\emph{``something that has two or three levels of parentheses \ldots I will try to break them apart into smaller pieces with more comments''}.

\section{RQ3 \& RQ4: What do Developers Find Difficult About Programming Regexes, and How do They Handle Those Difficulties?}
\label{section:Challenges}

% \FS{Consider clarifying what questions are we reporting answers from. I know that we asked specifically about difficulties in validating regexes, but what other questions are we taking these quotes from?
% I think it's from
% ``In your experience, what makes regular expression validation difficult (e.g., lack of tools to assist in validation)?''
% ``Can you tell us more about your thought process when you decide to reuse a regex vs writing it from scratch?'',
% ``Which of the following have you actually worried about in the past when copy-pasting a regex? How did you address it?'',
% ``Which of the following have you encountered in the past when copy-pasting a regex? How did you address it?''
% }

Intertwined with their decision-making process, participants mentioned many difficulties during regex programming.
As in the preceding section, we analyzed our survey responses and then used our interview phase to probe for additional details about the challenges that developers face and the ways in which they handle them.
%After analyzing these mentioned difficulties, we performed individual interviews to obtain further details about them, as well as to learn how developers handle them.
For clarity, in this section we accompany each challenge with the way(s) in which our participants described handling them\footnote{We chose our words carefully here. We refer to what participants described as ``handling problems'' because many participants were aware of limitations or gaps in their approaches. We discuss potential research directions to address some of their concerns in~\cref{section:Implications}.}.
\cref{table:Challenges} summarizes our findings.
As indicated in~\cref{table:Challenges}, the first two challenges we discuss were cross-cutting, spanning several stages of the regex programming process.
We then discuss two common challenges that our participants face when composing regexes, and two challenges that they face when validating regexes.

{
\renewcommand{\arraystretch}{1.4}
\begin{table*}
\small
\centering
%\raggedright
\caption{Developer-reported Difficulties when Programming Regexes and Handling Mechanisms for Them}
\begin{tabular}{lll}                                                                                                          
% \textbf{Stage} &\textbf{RQ3: Difficulty} & \textbf{RQ4: Handling Mechanism for Difficulty}\\ 
% \hline
% \multirow{2}{*}{\emph{Cross-cutting}} &\multirow{1}{*}{A) Understanding the Problem} & Read data, break down problem \\

%   &\multirow{1}{*}{B) Understanding the Regex} & Using tool support, breaking down regexes, adding documentation\\
%     \arrayrulecolor{black}\hline
    
% \multirow{2}{*}{\emph{Composing}} &\multirow{1}{*}{C) Searching for Reuse Candidates} & Decomposing the regex, searching for similar code, personal regex library\\

%   &\multirow{1}{*}{D) Non-intuitive Syntax} & Using tool support reading the regex documentation\\
%  \arrayrulecolor{black}\hline
%  \multirow{2}{*}{\emph{Validating a Regex}} &\multirow{1}{*}{E) Testing Edge Cases} & Test all available input\\

%  &\multirow{1}{*}{F) Testing Enough Inputs} & Work to find or generate more real cases, property testing\\
\toprule
\textbf{Stage}						&	\textbf{RQ3: Difficulty}							&	\textbf{RQ4: Handling Mechanism for Difficulty}	\\ 
\toprule
\multirow{5}{*}{Cross-cutting}		&	\multirow{2}{*}{A) Understanding the Problem}		&	Generalizing sample inputs										\\
									&														&	Breaking down the problem 								\\
\arrayrulecolor{gray}
\cline{2-3}
 									&	\multirow{3}{*}{B) Understanding the Regex}			&	Using tool support for visualization and highlighting								\\
									&														&	Breaking down regexes							\\
									&														&	Adding documentation							\\
\arrayrulecolor{black}
\hline
\multirow{5}{*}{Composing a Regex}	&	\multirow{3}{*}{C) Searching for Reuse Candidates}	&	Decomposing the regex							\\
									&														&	Searching for similar code						\\
									&														&	Personal regex library							\\
\arrayrulecolor{gray}
\cline{2-3}
 									&	\multirow{2}{*}{D) Non-intuitive Syntax}			&	Reading the regex documentation								\\
									&														&	Using tool support					\\
\arrayrulecolor{black}
\hline
\multirow{4}{*}{Validating a Regex}	&	\multirow{2}{*}{E) Testing Edge Cases}				&	Generating their own inputs						\\
                                                       &                                                                  & Testing all the available inputs           \\
\arrayrulecolor{gray}
\cline{2-3}
									&	\multirow{2}{*}{F) Testing Enough Inputs}			&	Request additional inputs from other stakeholders		\\
                  &                           & Property-based testing                \\
\arrayrulecolor{black}
\bottomrule
\end{tabular}

\label{table:Challenges}
\end{table*}

}

\subsection{Understanding the Problem}
\label{sec:difficultyUnderstandingProblem}

\subsubsection{Difficulty}
Multiple developers reported the difficulty of fully understanding the string problem that they are trying to solve, \eg
\emph{``The most difficult thing with regular expressions tends to be defining the problem''}.
% stages affected
Developers mentioned this difficulty affecting many of the stages of programming with regexes.
In particular, one participant mentioned understanding the problem as a difficult step in both writing and validating regexes with tests: 
\emph{``Having clear understanding of the task \ldots helps not only to write the tests but also to write/pick the regex''}.
This may be tied to the importance of a good set of inputs during the validation step, mentioned by other participants in~\cref{sec:RQ2-validating}.
%Apparently regex-style pattern matching can be ill-defined problem.
% \FS{Do we have quotes for this difficulty affecting decisions in problem assessment or finalizing?}

% related difficulties

% Handling challenge

\subsubsection{Handling}

% \FS{How do developers handle it?}
Developers primarily handled the difficulty of problem definition by \textbf{studying sample inputs} for patterns. %searching for patterns as their main method to understand a problem.
% \emph{``My first step is of course understanding what I'm actually trying to capture because in my line of work I tend to have anywhere from a hundred to 14,000 samples of what I'm looking for\ldots I tried to generalize what I'm looking at and then just craft the regular expression from there.''}
\emph{``I tried to generalize what I'm looking at and [then] craft the regular expression.''}
% \MM{I have at least one good quote for this but I need to find it}
When they could not understand the problem at a glance, developers discussed \textbf{breaking the problem into smaller pieces}.
% \emph{``I generally just have to get very methodical about what I am trying to do. I think about, here's my inputs, and these are the things in this input stream that I want to match, and then I ultimately want to transform this into an output string based on these things. I attack very much like I attack any programming problem, where I break it into manageable pieces.''} 
One interviewee described this as  \emph{``[getting] very methodical \ldots much like I attack any programming problem, where I break it into manageable pieces.''} 
%\DY{After reading the above two paragraphs, now I see that the difficulty we are discussing here is to define the language -- a set of strings to match and not to match.}

\subsection{Understanding the Regex}
\subsubsection{Difficulty}
% summary
Their opinions about regexes being ``self-documenting'' notwithstanding, many of our participants said that they perceive regexes as difficult to understand.
Several interviewees summed this up well, variously remarking that \emph{``The syntax of regular expressions is kind of terse''}, that \emph{``The regex syntax is cryptic''}, and simply referring to it as \emph{``Illegible gibberish''}.
% stages affected
Developers reported that \textbf{the difficulty of interpreting regex syntax as a pattern, \ie ``reading regexes'', impacted all stages of the regex programming process}.
The difficulty of understanding regexes is exacerbated by frequent \emph{``lack of comments/documentation, poor style''}, a point raised by many developers.

% validating - no documentation
% validating
%In relation to validation, 
%the ``terse'' syntax and frequent lack of built-in syntax highlighting in IDEs leaves developers struggling, saying things such as \emph{``Because the regex syntax is so cryptic, for more complicated regexes it is sometimes hard to know all the cases that should be tested.''}
%One developer simply stated \emph{``Illegible gibberish''} when asked about what they felt was difficult about regex validation. 
% Developers reported to be impacted by the readability of regexes 
% % when assessing how to solve the problem, 
% when composing them, 
% % deciding to use them, deciding to reuse them, when choosing a regex to use, when composing or modifying them, 
% when validating them, and when finalizing them. 
% \FS{Revisit this sentence}

% composing
%\DY{Well, the first sentence now talks about ``reuse'', trust, complexity. I think this is off the topic.}
%In relation to composition, often developers shy away from composing in favor of reusing because of the complex nature of regexes, stating things like: \emph{``I don't always trust myself to write complex ones myself''}.

% \MM{People use the word remember a fair amount in interviews they would struggle with remembering syntax which would make it hard to read later}
% \FS{Do we have quotes for this?}

% \FS{I couldn't find a quote for the difficulty of understanding the regex in relation to ``finalizing''}

% related difficulties

% Handling challenge
\subsubsection{Handling}

Developers reported two ways to handle this difficulty: using tools to improve their own regex comprehension, and documenting regexes to improve comprehension for others.
% handling it for current readers
The most common tools mentioned by participants were visualization aids, like graphical visualization and syntax highlighters.
The most praised among these were the \textbf{built-in syntax highlighting} for the JetBrains IDEs, \emph{``Jetbrains has my back - IDE syntax highlighting''}. 
% \FS{quote?}

% handling it for future readers
To improve regex comprehension for others, developers described their regex documentation strategies.
One common method was to \textbf{break regexes across multiple lines}, providing a comment about each individual part of the regex.
This is not a universally supported feature across all regex implementations, but was encouraged by participants when it was available. 
From a developer interview: \emph{``Hopefully, you're using a programming language where you could break it into multiple lines and comment those''}.

In addition, some developers encourage others to \textbf{document their regexes} when they come across them in code review.
In those cases, some developers push to include more detailed commenting, \eg \emph{``Put a plain language explanation in comments\ldots Have as many examples of matching and unmatching text as is appropriate in the comments''}

\subsection{Searching for Reuse Candidates}

In this and the following subsection, we introduce challenges specific to two stages of the regex programming process: composing and validating regexes.

\subsubsection{Difficulty}
In order to reuse a regex, developers first need to search for a regex that is worth reusing. 
% stages affected
Multiple participants lamented the difficulty of this process, which mostly affects the regex composition stage.

When our participants search for a regex to reuse, they usually try to leverage general search tactics with mixed success. 
In particular, \textbf{developers find it difficult to frame their search in a way that is understood by existing search tools} --- it is difficult to express their abstract string-matching problem as a plain-text query for a search engine.
For some tasks, the desired regex is easy to search, \eg ``email regex'', but developers often find it difficult to express the regex that they need.
From a developer interview: \emph{``It's hard to \ldots query the problem you're trying to solve. Sometimes it's so domain specific.''}

% related difficulties
In this case, developers face both the difficulty of understanding the problem itself (\cref{sec:difficultyUnderstandingProblem}) and the difficulty of articulating it for existing search engines.

% Handling challenge
\subsubsection{Handling}
Developers reported three mechanisms to handle this difficulty.

% decompose
Some developers choose to \textbf{decompose the regex} into smaller pieces that may be easier to search.
This is expressed well by one of the survey respondents: \emph{``If I can't find an existing regex that fits my need, I will start searching \ldots [for] pieces that will help me construct the final regex.''}

% search for code
Anther popular approach was an indirect search.
Developers commonly \textbf{search for code} that may use a relevant regex.
For example, this was described in an interview when the participant said:
\emph{``I would say intuition, but also sort of like code that's close by for sure. A file next to it or maybe it's an implementation of something that conforms in some interface.''}

% personal lists
Finally, some participants maintain \textbf{personally curated lists for regex reuse} that they will consult.
For example, one survey respondent mentioned that they would \emph{``refer to the regex section of my personal notebook''} when searching for reuse candidates.
This handling mechanism seems similar to having a personal library of utility functions that you copy into your codebase as needed.
% \FS{Discuss that this is interesting later in the discussion section.}
% This is an interesting phenomena that is not frequently seen in other kinds of code search or reuse behavior 

\subsection{Non-intuitive Syntax}
\subsubsection{Difficulty}

Developers frequently discussed dealing with the syntax of regexes as an obstacle, \eg
%``They're very non-intuitive.
\emph{``You have to remember what each symbol in that string means \ldots they're non-intuitive.''}
%that doesn't mean that anywhere but a regex''}
Another developer expressed a reliance on regex-specific reference charts:
\emph{``I need a little cheat sheet that has to outline what each symbol does.''}
%I don't use regular impressions to a point where it's useful for me to have it memorized. So I need like sheet to kind of see, what the star means''}
% \FS{Do we have quotes to support this point? Do we have quotes that point to the difficulty of composing more generally (not specifically in a different dialect)?}

% stages affected
% related difficulties

% Handling challenge
\subsubsection{Handling}

Developers relied on two handling mechanisms to address difficulties with syntax.

Unsurprisingly, developers often mentioned referring to their programming language's \textbf{regex documentation} to support their composition of regexes, \eg:
\emph{``\ldots If [searching for a regex to reuse] fails, I will start reading the regex documentation.''}

But another common mechanism is the usage of \textbf{tool support}.
\emph{``Anytime I am curious about a regex [I] go to regex101.com\ldots You type in your regex and some examples and it'll match or not match in real time and that's just useful.''}
% \FS{Do we have a quote for this?}
Developers also appreciated tools that incorporated documentation.
One participant noted that
% \emph{``[discussing regexr.com] there's a, bar on the side of super helpful. [You can] click on every sort of regular expression piece of syntax and at all it'll show you an example, a small example of what it does, how it works, and then there's a little brief summary. Even if I'm not using it to actually compose regular expressions, I'll go there and reference it.''}
\emph{``there's a bar on the side of [regexr.com] \ldots [You can] click on every sort of regular expression piece of syntax and it'll show you an example.''}
%\ldots Even if I'm not using it to actually compose regular expressions, I'll go there and reference it.''}
%This participants' advice for someone getting started with regexes was simply to ``use regexr.com''

\subsection{Testing Edge Cases}
\subsubsection{Difficulty}
Developers also found it difficult to identify \emph{corner cases} or \emph{edge cases}, terms colloquially used to refer to \emph{boundary values} \cite{Reid2002AnTesting}.
They often expressed the \emph{``Difficulty in [coming up with] corner case inputs and outputs''} and the fact that it is \emph{``Tough to imagine all edge cases to test''}.
% \FS{These quotes feel incomplete too.}

% stages affected
% related difficulties

\subsubsection{Handling}
Some developers handle this problem by generating their own input, while others rely on real-world input data from others.
For simple problems, developers say that they think through the problem space and generate their own input, but noted that this approach has its limits.
% \emph{``First of all, it would depend on the complexity of the regular expression. If the regular expression is one that is quote unquote simple enough that thinking about the entire scope of the input space, is actually feasible, all I really need to test is does the thing I matching occur at the beginning, middle or end of the string, matching have zero one or more. It's really the case where it really grows into a massively complex one that being able to thoroughly test it becomes a little less feasible.''}
\emph{``If the regular expression is \ldots simple enough that thinking about the entire scope of the input space is feasible \ldots It's really the case where it really grows into a massively complex one that [is problematic].''}

Other developers gained an understanding of their problem by reading data they had on hand, but they acknowledge that they therefore tend to \textbf{only address the edge cases that manifest in the available input data}.
%This leads them to simply test with all of the data they can to try to identify edge cases.
% \FS{Do we have anything for this?}
One participant remarked that \emph{``[I] look for everything that I can get from production\ldots that's my input \ldots that's my unit test.''}, and
went on to say \emph{``[but] unfortunately the input that I get can't be `universal'.''}.
% They discussed wishing they could fuzz their input at times but being unable to, \emph{``Sometimes even inputs from production unfortunately the input that I get can't be quote unquote universal. \ldots I can't fuzz input, but, it's the next best thing''}

\subsection{Testing Enough Inputs}
\subsubsection{Difficulty}
Similarly to testing edge cases, developers also reported on the difficulty of validating regexes with enough inputs.
Regexes are powerful and flexible tools, but in consequence can be very difficult to validate thoroughly.
A survey respondent summed up this idea well: 
\emph{``an infinite regression problem, to test a regex \ldots would require regexes''}.
In particular, developers find it difficult to come up with sets of sample inputs for testing that they would consider complete.
This was a frequent complaint: \emph{``insufficient sample inputs / unknown set of sample inputs''}
% \FS{This quote feels incomplete.}
% \MM{Still? I added some context that I thought might help}
% stages affected
% related difficulties

\subsubsection{Handling}
% \FS{Do we have anything for this?}
% \MM{Yes but I need to dig a little about what we can say specifically, right now my intuition is that either devs seek assistance from customers/testers to get more data or they just admit defeat}
Developers handled this through ad-hoc approaches to expand their collection of inputs.
One participant relied on the \textbf{expertise of other humans} to do this: 
\emph{``Testing literally every scenario is unfortunately not a realistic solution\ldots working with the QA and the clients to get a diverse set of real world documents''}.
Another participant said that they automatically generated additional input using \textbf{property-based tests}~\cite{propertyBasedTest}.

\section{RQ5: Are Developers Aware of Portability and Security (ReDoS) Risks when Programming Regexes?}
\label{section:Security}
As discussed in~\cref{sec:background}, developers encounter two risks when programming regexes: correctness concerns due to regex (non)-portability, and security concerns due to \REDOS.
In this section, we describe developers's awareness of these risks, as well as their handling mechanisms.
% \DY{I suggest flipping the following two sentences each other -- portability first and REDOS later.}

%Regex portability problems vary in how they manifest but the underlying issue is that many regex dialects exist across different regex implementations and are not necessarily compatible.
%\REDOS is an algorithmic complexity style security venerability that can result in a denial or service attack if a super-linear regex is exposed to malicious input.
% \FS{Why do we care about this? Motivate}

\begin{figure}
\centerline{\includegraphics[width=1.0\columnwidth]{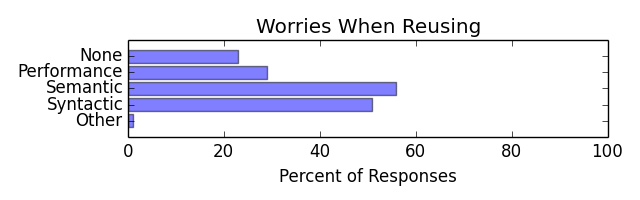}}
\caption{When asked what they worried about when reusing regexes, developers expressed a range of concerns, emphasizing semantic portability issues --- that a regex would not work as intended. Developers also worried less about performance issues, where a regex could slow down overall execution time, but this may be attributed to lack of awareness of performance vulnerabilities, as is discussed in ~\cref{subsection:ReDoS}
}.
\label{fig:MischaReuseWorry}
\end{figure}

\subsection{Portability Risks}
% \DY{Are we talking about porting a regex from one lang to another lang? Or even within the same lang as well?}
% \DY{I think here is the first time to introduce ``syntactic'', ``semantic'', and ``performance'' differences. Are they obvious? For performance, common case? worst-case?}
\subsubsection{Awareness}
% what we did
We asked developers whether they worry about a series of risks involved in regex reuse: syntactic, semantic, and performance differences when re-using regexes.
We report their answers in ~\cref{fig:MischaReuseWorry}.

% finding
\textbf{Developers who responded to the second survey worry about syntactic and semantic differences (over 50\%), and around a third (29\%) also worry about performance differences}.
Our interview participants provided more details about specific portability problems that they have resolved in the past.
They reported re-factoring regexes with unsupported features, which at times would blend semantic and syntactic differences, \eg \emph{``escape sequences \ldots vary across systems''} and \emph{``[Go regexes] don't support some of the constructs that are available in other languages''}.
% Dealing with mismatching feature support between languages is not only a tedious process, but also a difficult one, as we discussed in ~\cref{sec:crossCuttingIssues}.

% another finding
An interesting facet of~\cref{fig:MischaReuseWorry} is that a substantial portion of developers also reported not worrying about any of these reuse risks.
Our interviews shed some light into why many developers do not worry about regex reuse risks: \textbf{many are not aware that reuse carries portability risks}.
% \DY{Why are we not putting the following point in VII. RQ2. 1) Using regex vs. Alt.?}
In fact, some developers reported that they prefer to use regexes over other alternatives because of their (perceived) portability across languages.
One survey respondent stated \emph{``It is consistent across languages''}, and another said that the \emph{``same regex can be used across technologies/systems''}.
In concurrent work, we have explored this \textbf{misconception}~\cite{Davis2019WhyExpressions}.
% \MM{from James' survey}

% \DY{New paragraph?}
Furthermore, other respondents described the shock that they felt when they first learned of regex portability issues, \eg 
% \emph{``the most infuriating thing I have ran into when I was using this was the, the syntax differences. What support and what isn't. Because nobody really tells you that when you're first starting out with this stuff. 
\emph{``I certainly didn't know [before that incident] \ldots what the hell is that?''}
Ignorance of this issue exposes developers to correctness issues due to improper regex validation\footnote{For example, consider the participants that we described across several earlier sections, who reuse from a ``trusted source'' like Stack Overflow and do not validate carefully as a result.}.
We note that these assumptions can affect correctness whether or not regexes are being reused, simply as a result of an incorrect mental model for regex behavior.
%as a result making critical mistakes in the correctness of your regex or even possibly in its run time performance. 
% implications
% \FS{Briefly address ``so what?'' or refer to it in ``Implications''}

% reducing risk
\subsubsection{Handling}
% \FS{Address prevention methods separately for each risk}
Most developers handled missing feature support or blatant syntax differences by consulting language documentation and making a \textbf{translation}.
%This sentiment was expressed by a survey respondent describing how they ease their worries around reuse.  
For example, \emph{``for syntactic differences, I look at a regex cheatsheet and find the appropriate syntax for my environment''}.
%, then fix the regex.''}
% \MM{From easy worry survey}
If they understand the need for translation, developers do not find this translation difficult, though it can be \textbf{frustrating}.
In some cases, the need for regex translation can even influence larger project decisions.
One participant was considering migrating a project from one language to another, but decided against it: \emph{``Transitioning our common regular expressions \ldots kind of a headache.''}
%In an interview where a developer was discussing considering changing languages for a project:
%In fact regex translation was one of the factors that was considered when deciding to \textbf{not} change to this specific programming language. 
%\DY{I got a mixed feeling about this whole section (portability risks). For those who are aware of portability risks, this is not becoming the difficulties in searching reuse candidates, which we can discuss in VIII. RQ3 and RQ4 C. Searching.}
Other developers would not go to the effort to make a translation.
If they found that a regex reuse candidate would not work in their regex dialect, they would \textbf{start their reuse search process over} to find one that did not need a translation: 
\emph{``Sometimes I ported it. Sometimes I went looking for another.''}
% \MM{Ease worry response}
% \MM{From interview}
The primary way that most developers discussed dealing with further concerns was \textbf{testing} to confirm that the regex that they were reusing behaved in the way that they expected. 
\emph{``I run the regex against various tests to ensure it outputs as expected.''}
% \MM{ease worry codes}

% \begin{figure}
% \centerline{\includegraphics[width=1.0\columnwidth]{figs/MischaSurvey/redos.png}}
% \caption{Aggregate developers are overwhelmingly unaware of the possible security issues posed by regex usage}
% \label{fig:mischaREDOS}
% \end{figure}
% \begin{figure}
% \centerline{\includegraphics[width=1.0\columnwidth]{figs/JamesSurvey/AcmeRedos.png}}
% \caption{Only 40\% or ACME developers know what REDOS is.}
% \label{fig:JamesREDOS}
% \end{figure}

\subsection{ReDoS}
\subsubsection{Awareness}
\label{subsection:ReDoS}

% \FS{I don't think that we need figures for a yes/no answer. Simply report the percentages.}

Regexes open the somewhat obscure security concern of \REDOS.
\REDOS is fairly avoidable, however, if proper steps are taken to sanitize input and to not use super-linear regexes. 
Most importantly, the first step to avoiding the issue is being aware of the problem.
When asked if they knew what \REDOS was, developers were overwhelmingly unaware.
\textbf{Only \TotalPercKnowREDOS of all surveyed developers knew of the possible vulnerability.}
This is concerning, since the vulnerability is easy to introduce without noticing and it can slip through validation without being detected. 
We note that \REDOS is a vulnerability rather than an exploit, and is only relevant if the regex may match against malicious input. 
% This means that developers whose regexes process known or trusted input need not be concerned about it.
Nevertheless, we were surprised that the majority of participants were unaware of \REDOS.

\subsubsection{Handling}
Beyond their (occasional) awareness of this security risk, developers currently do little to combat performance issues and feel ill-equipped to identify performance issues leading to \REDOS in their regexes. 
When discussing performance issues and challenges in interviews, many developers said they only worry about regexes that introduce noticeable \emph{``lag''}: \emph{``I just wait and see if it becomes an issue.''}
%Some developers do mention \emph{``benchmarking''} as part of how they ease their worries around regex reuse but the far more popular course of action is much closer to run some tests and if it is not noticeable it is probably fine. \emph{`` I just wait and see if it becomes an issue.''}
% \emph{``Performance impact: assumed it'll appear in the profile if it actually cause performance problem that's worth fixing.''}
Only one interviewee referred to \REDOS (\emph{``catastrophic backtracking''}) thanks to a related feature in regex101.com.

Some developers held a misconception about worst-case regex performance issues, which may not manifest on typical input, but only on malicious input.
The feature in regex101.com is not a \REDOS detector, but rather a diagnosis tool suitable if the developer already knows about the worst-case input.

Part of these behaviors may be because developers \textbf{lack tools or knowledge} about solving regex performance problems.
For example, Davis \etal reported that input sanitization is an easy mitigation for many \REDOS vulnerabilities~\cite{Davis2018EcosystemREDOS}, but none of our participants mentioned this approach.
And when asked what is difficult about validation, one developer simply stated \emph{``Performance and security risks''}, but did not mention any of the tools tailored to this problem~\cite{Shen2018ReScueGeneticRegexChecker, Rathnayake2014rxxr2,Weideman2016REDOSAmbiguity,Wustholz2017Rexploiter}.
%Tools however do exist to detect super-linear regexes \cite{Shen2018ReScueGeneticRegexChecker, Rathnayake2014rxxr2,Weideman2016REDOSAmbiguity,Wustholz2017Rexploiter}.

%but it is a misconception that this would protect against \REDOS. The website would notify the user if the regex they were trying to work with was preforming catastrophic backtracking on the input they provided, not if it could be provided input that this would be the case for.

\section{Discussion and Implications}
\label{section:Implications}
Our findings have implications for many research directions.
% We discuss them below.

\subsection{Addressing String-Matching Problems}
\label{sec:implObservations}
Our observations motivate future studies of string-matching problems to understand them in more detail.

\subsubsection{Understanding String-matching Problems}
Developers specifically mentioned the difficulty of fully understanding string matching problems, which makes it hard to decide whether a regex is the best solution for the problem.
It may be beneficial to investigate in more depth how experts solve specific kinds of string-matching problems.
Understanding and naming categories of problems and their solutions would simplify the way that we describe and reflect about them --- much like it is done with design patterns in object-oriented programming \cite{Gamma1993DesignDesign}.
An example taxonomy member may take the shape of classifying all regexes that search for nested delimiters as a specific grouping.
Chapman and Stolee \cite{Chapman2016RegexUsageInPythonApps} proposed a classification of common regex solutions in terms of their textual similarity.
We propose a complementary research direction, by classifying string-matching problems.

\subsubsection{Diverse Solutions for String-matching Problems}
Furthermore, our participants mentioned that some string-matching problems are better addressed by non-regex solutions --- those that are either too simple or too complex.
It would be worth investigating what specific characteristics make string-matching problems fall into this category, and what kinds of solutions developers employ in those cases.
Some example alternative solutions may be: position-fetching, substring matching, anchor-splitting, or their combinations with simple regexes.
% CFGs
Another powerful mechanism that may be useful for this kind of problems are context-free grammars --- \eg using tools like ANTLR \cite{parr2013definitive}, which may also be easier to understand and debug.
It is possible that an additional level of abstraction or more powerful language with built-in string functions could solve some of the challenges that developers face with regexes.

\subsection{Assessing Regexes}
\label{sec:implMetrics}
% observed need
We found that developers need to assess various qualities of regular expressions.
Our findings for RQ2 and RQ4 explain how developers use various factors of \textbf{complexity} and \textbf{quality} to make decisions and overcome difficulties.
For example, developers considered both regex complexity and quality as important factors to decide which regexes to reuse.
% state of the art limitations
Such estimations of regex complexity and quality are normally performed manually, by simply looking at the regex.
% Many of these measurements are based on easy-to-see characteristics --- like length and features used.

% research direction
\subsubsection{Regex Metrics}
These findings highlight the value of developing regex metrics to automatically measure the quality and complexity of regexes in a way that will help developers make decisions when programming regexes.
For source code, metrics already exist to capture some of its complexity and quality, \eg cyclomatic complexity \cite{McCabe1976AMeasure}.
% However, these metrics have not been adapted to regular expressions in particular --- with the exception of Wang \etal's method \cite{Wang2018RegexTestCoverage} of measuring test coverage for regexes as a measure of quality.
We pose that regex metrics would have a strong impact in the productivity of developers when programming regexes, since most of the decisions that they make consider some metric.
We elaborate further on our envisioned usage of regex metrics in the following sections.

\subsection{Automated Support for Reusing Regexes}
\label{sec:implReuse}
% observation
Our participants highlighted multiple difficulties in regex reuse, such as defining the problem for a search engine query and reusing across regex dialects.
They also pointed out the characteristics that they valued when reusing, as well as the interesting practice of keeping regex lists for reuse.
These findings open multiple opportunities for research.

\subsubsection{Semantic Regex Search}
% state of the art
Currently, developers find it difficult to use search engines to look for regexes to reuse, since it is hard to express their string-matching problem in a few words --- particularly considering that the problems themselves are hard to understand (see \cref{sec:difficultyUnderstandingProblem}).

% idea
Thus, developers would benefit from a search engine that allowed them to express string-matching problems in a domain-specific way, \ie semantic regex search.
Such a semantic regex engine could be more useful to developers by taking inputs that are regex-specific.
For example, these could be: 
(a) a list of inputs that match or non-match the regex that they need,
(b) a regex that resembles the regex that they need
(c) a code context in which the regex would be used.
The approaches for processing these inputs would vary, but we pose that such a search engine would make it easier for developers to express the problem that they are trying to solve.

\subsubsection{Regex Repositories}
% idea
Since developers mentioned keeping their own lists of regexes for future reuse, it may be useful to empower them in that practice.
Regex repositories would strongly complement semantic regex search, particularly if the stored regexes contain additional metadata, such as: the category of string-matching problem that they solve (possibly from Chapman and Stolee's \cite{Chapman2016RegexUsageInPythonApps} proposed classification of common regex solutions), portability or performance problems, or other indicators of quality via regex metrics or user ratings.
One regex repository does currently exist (RegExLib \cite{RegexRepository_RegexLib}), but it encompasses a very narrow set of facets in which to perform search, and is also relatively small given the relative size of other existing software datasets. 
% with annotations
% with design patterns

\subsubsection{Metrics-based Regex Ranking}
% idea
Regex metrics would complement semantic regex search by allowing the ranking of results according to various metrics.
Developers expressed that, in most regex-reuse cases, they valued low complexity --- \eg short length, few features used --- and high quality --- \eg coming from a reliable source and being better tested. 

% state of the art

\subsubsection{Regex Dialect Translation}
% idea
Finally, developers also expressed the difficulty of reusing regexes that were created in a different dialect.
Ideally, a regex search engine should also include mechanisms to understand the regex dialect for which a regex was created, as well as mechanisms to refactor it for the dialect in which it will be used.

% state of the art

\subsection{Automated Support for Composing Regexes}
\label{sec:implComposing}
% observation
We also identified developers' difficulties with composing regexes --- \eg dealing with difficult syntax that is hard to remember,
and many qualities that developers value in regexes --- \eg short length and reduced feature usage.
% state of the art
These findings motivate many avenues for research.

% idea
\subsubsection{Live Support for Regex Composition}
% for better metrics
% for better breakdown
% to transform between match too much vs. match 
% too little
Now that we better understand some of the characteristics that developers value in regexes (see \cref{sec:implMetrics}), we could develop assistants to support developers in composing regexes with those desirable attributes.
Such assistants could help developers to, \eg break down their regexes, use fewer advanced features, or decide between matching too match or too little.

\subsubsection{Regex Refactoring}
% search first
The same information about desirable qualities in regexes could be applied to develop regex refactorings that would improve the quality of regexes.
Regex refactoring could be applied on demand or automatically over a whole code base.
Refactoring is another example of a highly-valued concept in object oriented programming that could highly benefit regex programming.
Chapman and Stolee \cite{Chapman2016RegexUsageInPythonApps} already proposed the idea of regex refactoring.
Our findings throw more light into what kinds of refactorings would be desirable for developers.

\subsubsection{Automatic Regex Composition}
Another interesting research direction would be the full automation of regex composition.
Some existing work has made advances in this direction by composing regexes from examples of matching and non-matching input \cite{txt2re, RegexMagic, Bartoli2016InferenceExamples, Bartoli2014PlayingProgramming}
% \FS{I remember this from earlier conversations, but I don't have a citation.}
A different approach to automate regex composition may feed the algorithm with partial regexes that solve pieces of the problem --- since developers seem to be already decomposing the problem in their manual composition efforts.
% based on matching/non-matching input
% based on example seed regex / regex pieces

\subsection{Automated Support for Validating Regexes}
\label{sec:implValidating}
% observation
Developers mentioned the difficulty of validating regexes, particularly in testing edge cases and in deciding whether they tested enough input cases.
% state of the art
They currently handle such difficulties manually, by testing all the input that they have available.
These difficulties motivate research in at least two directions.

% idea
\subsubsection{Regex Input Generation for Humans}
Many tools have been proposed to generate input for testing regular expressions \cite{Larson2016GeneratingExpressions,Veanes2010Rex,Arcaini2017MutRex,Shen2018ReScueGeneticRegexChecker,Moller2010Brics}.
Surprisingly, none of our studied developers mentioned using these tools for input generation.

We realize that further study would be necessary to understand the extent to which these tools are adopted.
However, we believe that further research is motivated to study what developers consider \textit{good} or \textit{relevant} input, as well as to evaluate regex input generators when their output is consumed and judged by humans.

\subsubsection{Boundary-value Analysis for Regexes}
Methods like boundary-value analysis and equivalence partition \cite{Reid2002AnTesting} help software developers define and test \emph{edge cases} and are widely-known.
Thus, further research is motivated to adapt these techniques to regexes in particular or to develop other methods to support developers in defining boundary values for regexes.

% \FS{Another interesting point: All these complaints seem focused on white-box testing.}

\subsection{Automated Support for Documenting Regexes}
\label{sec:implDocumenting}
% observation
Some of our participants felt that regexes should be self-documenting, others thought that documenting regexes is necessary, and others thought that the decision depends on the complexity of the regex.
In addition to this, we also observed the things that developers value in regex documentation: breaking down the regex into pieces, documenting each piece, and the inclusion of both matching and non-matching input, as well as a plain description of what the regex does.

\subsubsection{Automatic Regex Documentation}
These findings motivate research to automatically document regexes, since many of the pieces of information that developers value could potentially be generated automatically.
Machine learning techniques could be developed to automatically break down regexes by \emph{learning} from examples.
Also, regex input generation techniques (see \cref{sec:implValidating}) could be adapted to generating few inputs that would be highly relevant for humans to understand the regex, and that also covered matching and non-matching input.

\subsection{Understanding Regexes}
\label{sec:implUndersRegex}
% observation
Many developers mentioned the \emph{terseness} and \emph{cryptic-ness} of the syntax of regexes, and how that makes them very difficult to understand.
% state of the art

% idea
\subsubsection{Novel Regex Syntax for Comprehension}
We believe that this widely-held sentiment calls for further research into new syntax for regexes to make them easy for humans to understand.
Since developers are already breaking down regexes in multiple lines, that may give space for a more verbose syntax that could be more easily understood.

\subsection{Regex Education}
Developers noted that regexes can be difficult to learn, but also consistently noted that they are a very helpful tool.
Our findings will help educators to teach best practices in programming with regexes, as well as specific difficulties to anticipate in the process.
\subsubsection{Learning Regexes} This also gives a clear incentive to investigating further how developers first learn regexes, as well as the common pitfalls that beginners face in particular.
% What are common pitfalls in very early understanding? 
This research brought to light some common mistakes that professional developers can make, such as perceived portability.
A similar study focused on developers learning regexes would provide more comprehensive insights into the early stages of understanding. 
\subsubsection{\REDOS Awareness}
Most developers do not know about \REDOS.
We believe that this vulnerability is easily overlooked, since regexes may seem \emph{harmless} to many developers.
Thus, we encourage educators and professionals to disseminate the mechanics of this vulnerability to better prevent problems associated with it.

\section{Threats to Validity}
\textbf{Construct Validity}:
In order to pose questions about the regex process to developers, we first outlined a general process based on our understanding of general software engineering practice. 
This may have limited the extent to which participants reflected on their own distinct decision-making factors or difficulties.
An example of findings that we may not have been able to observe are those related to code-review practices when regexes are involved, since we did not ask about code review explicitly.
To counteract this risk, we used open-ended questions, which enabled our participants to report on difficulties and decisions that we did not anticipate, \eg matching too much vs. matching too little.
Still, an observational study of our studied phenomena, based on contextual inquiry, may uncover a broader set of findings.

\textbf{Internal Validity}:
The researchers on this study manually analyzed the data by reading, summarizing, categorizing and discussing the contents of developer surveys and interviews.
This has the potential to introduce bias at various levels, and may also limit reproducibility, since other individuals may interpret the same information in different ways. 
This is a known limitation of qualitative research \cite{GolafshaniTheResearch}.
We strived to reduce the impact of this limitation by corroborating and discussing our findings across authors and across the two developer populations. 

\textbf{External Validity}:
We sampled hundreds of developers across two different populations, but this is still a small portion of all software engineers and populations.
In the second survey, we used snowball sampling for part of the sample, starting with professional contacts of the authors, biasing towards people known by the authors and at companies where the authors have worked.
Though this may compound some bias, we believe that our snowball sampling allowed us to contact both a large and wide range of developers, and the information that these participants provided was similar to the one provided by participants from our public recruitment approaches.

\section{Conclusion}
Regexes are powerful tools that developers find valuable.
Regexes are also hard to work with.
We identify six difficulties that developers face when using regexes and multiple handling mechanisms that they employ to deal with them.
% The way that developers cope leaves much room for improvement.
Developers are also mostly unaware of risks that they take when using regexes --- under 40\% of our participants were aware of security vulnerabilities associated with regex usage.
We propose many lines of research and new ways to support developers when working with regexes.

\section{Replication}
Our survey instruments and interview protocols are available for replication at \url{http://doi.org/10.5281/zenodo.3424069} \cite{louis_g_michael_iv_2019_3424069}.
% \FS{Can we say here again that we only analyzed the open questions in our instrument? It may reduce confusion. Can we add this to the PDFs themselves too?}

\balance

\bibliography{WebLinks.bib,personal.bib,references.bib,LinguaFranca.bib,references2.bib}

\begin{thebibliography}{10}

\bibitem{HackerNews}
Hacker news.
\newblock \url{https://news.ycombinator.com/}.

\bibitem{Reddit}
Reddit.
\newblock \url{https://www.reddit.com/}.

\bibitem{RegexRepository_RegexLib}
Regular expression library.
\newblock
  \url{https://web.archive.org/web/20180920164647/http://regexlib.com/}.

\bibitem{adolph2011using}
S.~Adolph, W.~Hall, and P.~Kruchten.
\newblock Using grounded theory to study the experience of software
  development.
\newblock {\em Empirical Software Engineering}, 16(4):487--513, 2011.

\bibitem{Arcaini2017MutRex}
P.~Arcaini, A.~Gargantini, and E.~Riccobene.
\newblock {MutRex: A Mutation-Based Generator of Fault Detecting Strings for
  Regular Expressions}.
\newblock In {\em International Conference on Software Testing, Verification
  and Validation Workshops (ICSTW)}, 2017.

\bibitem{Bacchelli2013ExpectationsReview}
{Bacchelli} and {Bird}.
\newblock {Expectations, Outcomes, and Challenges of Modern Code Review}.
\newblock In {\em International Conference on Software Engineering}, pages
  712--721, 2013.

\bibitem{Bartoli2014PlayingProgramming}
A.~Bartoli, A.~De~Lorenzo, E.~Medvet, and F.~Tarlao.
\newblock {Playing regex golf with genetic programming}.
\newblock pages 1063--1070. Association for Computing Machinery (ACM), 7 2014.

\bibitem{Bartoli2016InferenceExamples}
A.~Bartoli, A.~De~Lorenzo, E.~Medvet, and F.~Tarlao.
\newblock {Inference of Regular Expressions for Text Extraction from Examples}.
\newblock {\em IEEE Transactions on Knowledge and Data Engineering},
  28(5):1217--1230, 5 2016.

\bibitem{Biernacki1981SnowballSampling}
P.~Biernacki and D.~Waldorf.
\newblock {Snowball Sampling: Problems and Techniques of Chain Referral
  Sampling}.
\newblock {\em Sociological Methods {\&} Research}, 10(2):141--163, 11 1981.

\bibitem{Breu2010InformationReports}
S.~Breu, R.~Premraj, J.~Sillito, and T.~Zimmermann.
\newblock {Information needs in bug reports}.
\newblock In {\em Proceedings of the 2010 ACM conference on Computer supported
  cooperative work - CSCW '10}, page 301, New York, New York, USA, 2010. ACM
  Press.

\bibitem{Buse2012InformationAnalytics}
R.~P.~L. Buse and T.~Zimmermann.
\newblock {Information Needs for Software Development Analytics}.
\newblock In {\em Proceedings of the 34th International Conference on Software
  Engineering}, pages 987--996, Zurich, Switzerland, 2012. IEEE.

\bibitem{Chapman2016ExploringPython}
C.~Chapman and K.~T. Stolee.
\newblock {Exploring regular expression usage and context in Python}.
\newblock In {\em Proceedings of the 25th International Symposium on Software
  Testing and Analysis - ISSTA 2016}, pages 282--293, New York, New York, USA,
  2016. ACM Press.

\bibitem{Chapman2016RegexUsageInPythonApps}
C.~Chapman and K.~T. Stolee.
\newblock {Exploring regular expression usage and context in Python}.
\newblock {\em International Symposium on Software Testing and Analysis
  (ISSTA)}, 2016.

\bibitem{Chapman2017RegexComprehension}
C.~Chapman, P.~Wang, and K.~T. Stolee.
\newblock {Exploring Regular Expression Comprehension}.
\newblock In {\em Automated Software Engineering (ASE)}, 2017.

\bibitem{Cox2007RegexAlgorithms}
R.~Cox.
\newblock {Regular Expression Matching Can Be Simple And Fast (but is slow in
  Java, Perl, PHP, Python, Ruby, ...)}, 2007.

\bibitem{creswell2017research}
J.~W. Creswell and J.~D. Creswell.
\newblock {\em Research design: Qualitative, quantitative, and mixed methods
  approaches}.
\newblock Sage publications, 2017.

\bibitem{Crosby2003REDOS}
S.~Crosby.
\newblock {Denial of service through regular expressions}.
\newblock {\em USENIX Security work in progress report}, 2003.

\bibitem{Crosby2003AlgorithmicComplexityAttacks}
S.~A. Crosby and D.~S. Wallach.
\newblock {Denial of Service via Algorithmic Complexity Attacks}.
\newblock In {\em USENIX Security}, 2003.

\bibitem{Davis2018EcosystemREDOS}
J.~C. Davis, C.~A. Coghlan, F.~Servant, and D.~Lee.
\newblock {The Impact of Regular Expression Denial of Service (ReDoS) in
  Practice: an Empirical Study at the Ecosystem Scale}.
\newblock In {\em The ACM Joint European Software Engineering Conference and
  Symposium on the Foundations of Software Engineering (ESEC/FSE)}, 2018.

\bibitem{Davis2019WhyExpressions}
J.~C. Davis, L.~G. Michael~IV, C.~A. Coghlan, F.~Servant, and D.~Lee.
\newblock {Why aren’t regular expressions a lingua franca? an empirical study
  on the re-use and portability of regular expressions}.
\newblock In {\em Proceedings of the 2019 27th ACM Joint Meeting on European
  Software Engineering Conference and Symposium on the Foundations of Software
  Engineering - ESEC/FSE 2019}, pages 443--454, New York, New York, USA, 2019.
  ACM Press.

\bibitem{davis2019generalizability}
J.~C. Davis, D.~Moyer, A.~Kazerouni, and D.~Lee.
\newblock Testing regex generalizability and its implications: A large-scale
  many-language measurement study.
\newblock In {\em ACM International Conference on Automated Software
  Engineering (ASE)}. ACM, 2019.

\bibitem{Davis2018NodeCure}
J.~C. Davis, E.~R. Williamson, and D.~Lee.
\newblock {A Sense of Time for JavaScript and Node.js: First-Class Timeouts as
  a Cure for Event Handler Poisoning}.
\newblock In {\em USENIX Security Symposium (USENIX Security)}, 2018.

\bibitem{txt2re}
M.~J. Ennis.
\newblock txt2re.
\newblock \url{http://www.txt2re.com/}, 2006.

\bibitem{Ernst2015MeasureDebt}
N.~A. Ernst, S.~Bellomo, I.~Ozkaya, R.~L. Nord, and I.~Gorton.
\newblock {Measure it? Manage it? Ignore it? software practitioners and
  technical debt}.
\newblock pages 50--60. Association for Computing Machinery (ACM), 8 2015.

\bibitem{Fannoun2019TowardsPractice}
S.~Fannoun and J.~Kerins.
\newblock {Towards organisational learning enhancement: assessing software
  engineering practice}.
\newblock {\em Learning Organization}, 26(1):44--59, 1 2019.

\bibitem{friedl2006mastering}
J.~E. Friedl.
\newblock {\em Mastering regular expressions}.
\newblock " O'Reilly Media, Inc.", 2006.

\bibitem{Gamma1993DesignDesign}
E.~Gamma, , R.~Helm, , R.~Johnson, , and J.~Vlissides.
\newblock {Design Patterns: Abstraction and Reuse of Object-Oriented Design}.
\newblock In {\em ECOOP' 93 --- Object-Oriented Programming}, pages 406--431.
  Springer Berlin Heidelberg, 1993.

\bibitem{GolafshaniTheResearch}
N.~Golafshani.
\newblock {The Qualitative Report Understanding Reliability and Validity in
  Qualitative Research}.
\newblock Technical report.

\bibitem{Gousios2015WorkPerspective}
G.~Gousios, A.~Zaidman, M.~A. Storey, and A.~Van~Deursen.
\newblock {Work practices and challenges in pull-based development: The
  integrator's perspective}.
\newblock In {\em Proceedings - International Conference on Software
  Engineering}, volume~1, pages 358--368. IEEE Computer Society, 8 2015.

\bibitem{Hodovan2010RegularWeb}
R.~Hodov{\'{a}}n, Z.~Herczeg, and {\'{A}}.~Kiss.
\newblock {Regular expressions on the web}.
\newblock In {\em International Symposium on Web Systems Evolution (WSE)},
  2010.

\bibitem{louis_g_michael_iv_2019_3424069}
L.~G.~M. IV, J.~Donohue, J.~C. Davis, D.~Lee, and F.~Servant.
\newblock {Replication package for "Regexes are Hard: Decision-making,
  Difficulties, and Risks in Programming Regular Expressions"}, Sept. 2019.

\bibitem{Kitchenham2008PersonalSurveys}
B.~A. Kitchenham and S.~L. Pfleeger.
\newblock {Personal opinion surveys}.
\newblock In {\em Guide to Advanced Empirical Software Engineering}. 2008.

\bibitem{Ko2007InformationTeams}
A.~J. Ko, R.~DeLine, and G.~Venolia.
\newblock {Information needs in collocated software development teams}.
\newblock In {\em Proceedings - International Conference on Software
  Engineering}, pages 344--353, 2007.

\bibitem{l2001qualitative}
B.~L~BERG.
\newblock Qualitative research methods for the social sciences.
\newblock 2001.

\bibitem{Larson2018AutomaticCheckingOfRegexes}
E.~Larson.
\newblock {Automatic Checking of Regular Expressions}.
\newblock In {\em Source Code Analysis and Manipulation (SCAM)}, 2018.

\bibitem{Larson2016GeneratingExpressions}
E.~Larson and A.~Kirk.
\newblock {Generating Evil Test Strings for Regular Expressions}.
\newblock In {\em Proceedings - 2016 IEEE International Conference on Software
  Testing, Verification and Validation, ICST 2016}, pages 309--319. Institute
  of Electrical and Electronics Engineers Inc., 7 2016.

\bibitem{Li2015WhatEngineer}
P.~L. Li, A.~J. Ko, and J.~Zhu.
\newblock {What makes a great software engineer?}
\newblock In {\em Proceedings - International Conference on Software
  Engineering}, volume~1, pages 700--710. IEEE Computer Society, 8 2015.

\bibitem{lindlof2017qualitative}
T.~R. Lindlof and B.~C. Taylor.
\newblock {\em Qualitative communication research methods}.
\newblock Sage publications, 2017.

\bibitem{RegexMagic}
J.~G. S.~C. Ltd.
\newblock Regexmagic.
\newblock \url{https://www.regexmagic.com/autogenerate.html}, 2014.

\bibitem{propertyBasedTest}
D.~R. MacIver.
\newblock What is property based testing?
\newblock
  \url{https://hypothesis.works/articles/what-is-property-based-testing/}.

\bibitem{McCabe1976AMeasure}
T.~McCabe.
\newblock {A Complexity Measure}.
\newblock {\em IEEE Transactions on Software Engineering}, SE-2(4):308--320, 12
  1976.

\bibitem{Moller2010Brics}
A.~M{\o}ller.
\newblock dk. brics. automaton--finite-state automata and regular expressions
  for java, 2010, 2010.

\bibitem{parr2013definitive}
T.~Parr.
\newblock {\em The definitive ANTLR 4 reference}.
\newblock Pragmatic Bookshelf, 2013.

\bibitem{Pressman2010SoftwareApproachb}
R.~Pressman.
\newblock {Software Engineering: A Practitioner's Approach}.
\newblock chapter Process Models, pages 30--64. McGraw-Hill, seventh edition
  edition, 2010.

\bibitem{Rathnayake2014rxxr2}
A.~Rathnayake and H.~Thielecke.
\newblock {Static Analysis for Regular Expression Exponential Runtime via
  Substructural Logics}.
\newblock Technical report, 2014.

\bibitem{Reid2002AnTesting}
S.~Reid.
\newblock {An empirical analysis of equivalence partitioning, boundary value
  analysis and random testing}.
\newblock pages 64--73. Institute of Electrical and Electronics Engineers
  (IEEE), 11 2002.

\bibitem{Sadler2010ResearchStrategy}
G.~R. Sadler, H.-C. Lee, R.~S.-H. Lim, and J.~Fullerton.
\newblock {Research Article: Recruitment of hard-to-reach population subgroups
  via adaptations of the snowball sampling strategy}.
\newblock {\em Nursing {\&} Health Sciences}, 12(3):369--374, 9 2010.

\bibitem{Shen2018ReScueGeneticRegexChecker}
Y.~Shen, Y.~Jiang, C.~Xu, P.~Yu, X.~Ma, and J.~Lu.
\newblock {ReScue: Crafting Regular Expression DoS Attacks}.
\newblock In {\em Automated Software Engineering (ASE)}, 2018.

\bibitem{Siegmund2014MeasuringExperience}
J.~Siegmund, C.~K{\"{a}}stner, J.~Liebig, S.~Apel, and S.~Hanenberg.
\newblock {Measuring and modeling programming experience}.
\newblock {\em Empirical Software Engineering}, 19(5):1299--1334, 10 2014.

\bibitem{Spishak2012AExpressions}
E.~Spishak, W.~Dietl, and M.~D. Ernst.
\newblock {A type system for regular expressions}.
\newblock pages 20--26. Association for Computing Machinery (ACM), 7 2012.

\bibitem{Staicu2018REDOS}
C.-A. Staicu and M.~Pradel.
\newblock {Freezing the Web: A Study of ReDoS Vulnerabilities in
  JavaScript-based Web Servers}.
\newblock In {\em USENIX Security Symposium (USENIX Security)}, 2018.

\bibitem{SublimeSearchAndReplace}
S.~Team.
\newblock Sublime search and replace.
\newblock
  \url{http://docs.sublimetext.info/en/latest/search_and_replace/search_and_replace_overview.html}.

\bibitem{VSCodeBasicEditing}
V.~S.~C. Team.
\newblock Visual studio code - basic editing.
\newblock \url{https://code.visualstudio.com/docs/editor/codebasics}.

\bibitem{Veanes2010Rex}
M.~Veanes, P.~De~Halleux, and N.~Tillmann.
\newblock {Rex: Symbolic regular expression explorer}.
\newblock {\em International Conference on Software Testing, Verification and
  Validation (ICST)}, 2010.

\bibitem{WangExploringEvolution}
P.~Wang, G.~R. Bai, and K.~T. Stolee.
\newblock {Exploring Regular Expression Evolution}.
\newblock Technical report.

\bibitem{Wang2019ExploringEvolution}
P.~Wang, G.~R. Bai, and K.~T. Stolee.
\newblock {Exploring Regular Expression Evolution}.
\newblock In {\em Software Analysis, Evolution, and Reengineering (SANER)},
  2019.

\bibitem{Wang2018RegexTestCoverage}
P.~Wang and K.~T. Stolee.
\newblock {How well are regular expressions tested in the wild?}
\newblock In {\em Foundations of Software Engineering (FSE)}, 2018.

\bibitem{Weideman2016REDOSAmbiguity}
N.~Weideman, B.~van~der Merwe, M.~Berglund, and B.~Watson.
\newblock {Analyzing matching time behavior of backtracking regular expression
  matchers by using ambiguity of NFA}.
\newblock In {\em Lecture Notes in Computer Science (including subseries
  Lecture Notes in Artificial Intelligence and Lecture Notes in
  Bioinformatics)}, volume 9705, pages 322--334, 2016.

\bibitem{weiss1995learning}
R.~S. Weiss.
\newblock {\em Learning from strangers: The art and method of qualitative
  interview studies}.
\newblock Simon and Schuster, 1995.

\bibitem{Wustholz2017Rexploiter}
V.~Wustholz, O.~Olivo, M.~J.~H. Heule, and I.~Dillig.
\newblock {Static Detection of DoS Vulnerabilities in Programs that use Regular
  Expressions}.
\newblock In {\em International Conference on Tools and Algorithms for the
  Construction and Analysis of Systems (TACAS)}, 2017.

\end{thebibliography}
\bibliographystyle{abbrv}

\end{document}